\documentclass[paper]{JHEP3}
\usepackage{amsmath}
\usepackage{amsfonts}
\usepackage{amssymb}
\usepackage{graphicx}
\usepackage{tabularx}
\usepackage{array}
\usepackage{cite}
\makeatletter
\def\slash#1{{\mathpalette\c@ncel{#1}}} 
\preprint{Cavendish--HEP--09/12}
\title{Charge and mass effects on the evaporation of higher-dimensional rotating black holes}
\author{Marco O.P.\ Sampaio\\ 
  Cavendish Laboratory, University of Cambridge,
  J.J.\ Thomson Avenue, Cambridge CB3 0HE, U.K.\\ E-mail: \email{sampaio@hep.phy.cam.ac.uk}}

\abstract{To study the dynamics of discharge of a brane black hole in TeV gravity scenarios, we obtain the approximate electromagnetic field due to the charged black hole, by solving Maxwell's equations perturbatively on the brane. In addition, arguments are given for brane metric corrections due to backreaction. We couple brane scalar and brane fermion fields with non-zero mass and charge to the background, and study the Hawking radiation process using well known low energy approximations as well as a WKB approximation in the high energy limit. We argue that contrary to common claims, the initial evaporation is not dominated by fast Schwinger discharge.}


\begin{document}

\section{Introduction}
Recent proposals for extending the Standard Model of particle physics (SM) have suggested the existence of extra spatial dimensions as a solution to the hierarchy problem~\cite{Antoniadis:1990ew,Arkani-Hamed:1998rs,Antoniadis:1998ig,Arkani-Hamed:1998nn,Randall:1999vf,Randall:1999ee}. Such scenarios lower the strong gravity scale to 1~TeV and allow for the production of black holes at high energies~\cite{Giddings:2001bu,Dimopoulos:2001hw,Argyres:1998qn,Eardley:2002re}, such as the proton-proton collisions planned at the Large Hadron Collider (LHC). For consistency with current experimental observations~\cite{kapner:021101,tu:201101,Satheeshkumar:2008fb,Hannestad:2003yd}, the class of allowed models is constrained. Examples are the number of extra dimensions $n\geq 3$ and the fundamental Planck mass $M_{4+n}$. Another important constraint is the need to confine SM fields on thin branes to avoid bounds from electroweak precision observables~\cite{Flacke:2005hb} and fast proton decay~\cite{ArkaniHamed:1999dc,ArkaniHamed:1999za}. 
Then the SM fields generated by the charges trapped inside the hypothetical black holes will be confined to the brane, rather than spreading into the bulk.
Thus their influence in the Hawking decay will differ from the non-confined case. 

The study of black hole evaporation is also interesting as a theoretical arena in which to develop understanding of quantum field theory in curved space-time~\cite{Hawking:2005kf} and its possible extensions~\cite{Rovelli:2008zza,Green:1987sp}. For Cosmology, the evaporation of small primordial black holes could be relevant if they were produced after the Big Bang~\cite{Page:1976wx}. Finally in Astrophysics the tools employed to study Hawking radiation may be applied to issues such as black hole stability and scattering of waves around black holes~\cite{Regge:1957td,Teukolsky:1973ha,Press:1973zz,Teukolsky:1974yv,Dolan:2008kf}.   

Phenomenologically, the study of Black Hole (BH) events in high energy collisions has evolved through the development of event generators~\cite{Dimopoulos:2001hw,Harris:2003db,Cavaglia:2006uk,Dai:2007ki,Frost:2009cf}. The latter incorporate results from Hawking evaporation to simulate the decay phase (i.e. the evaporation) together with models for production.
In this paper, we focus on the evaporation of $(4+n)$-dimensional black holes on the brane. Several detailed studies appeared recently using analytical~\cite{Creek:2007sy,Creek:2007tw,Creek:2007pw,Creek:2006ia,Kanti:2002nr,Kanti:2002ge,Frolov:2002xf} and numerical tools~\cite{Ida:2002ez,Harris:2003eg,Harris:2005jx,Ida:2005ax,Duffy:2005ns,Casals:2005sa,Cardoso:2005vb,Cardoso:2005mh,Ida:2006tf,Casals:2006xp,Casals:2008pq}. They focused mainly on massive rotating black holes, so charge has been largely neglected. The usual motivation to start by neglecting charge relies on the claim that the black hole quickly discharges through Schwinger emission in the first stages of evaporation~\cite{Giddings:2001bu,Gibbons:1975kk,Carter:1974yx,Page:1977um}. We present detailed arguments to show that Schwinger emission alone does not suffice to discharge the black hole. This is due to the strengthening of the gravitational field compared to the electromagnetic field in TeV gravity scenarios which is in contrast with the weakness of gravity compared to electromagnetism in 4-dimensional Einstein-Maxwell theory. Thus the usual results in four dimensions, which favour strong Schwinger evaporation~\cite{Gibbons:1975kk,Carter:1974yx,Page:1977um}, do not hold.     

The structure of the paper is the following: In Sec.~\ref{sec:background} we construct the approximate Maxwell field of an electrically charged $(4+n)$-dimensional brane black hole starting with a background projected Myers-Perry metric with one angular momentum on the brane. In Sect.~\ref{sec:backreact}, we comment on backreaction and propose an effective metric to incorporate the effect of brane charge on the metric. Sect.~\ref{sec:OrdersOfMagnitude} is devoted to a detailed matching of the relevant couplings in the classical limit, taking into account the underlying assumptions of large extra dimensions models. Some remarks are made on the relative strengths of forces. In the beginning of Sect.~\ref{sec:Evaporation} the coupling of matter fields to the background is briefly presented. In Sec.~\ref{sec:HawkingRad} we summarize the relevant formulas from Hawking radiation which apply to our case, as well as a direct estimate from Schwinger's formula, indicating that Schwinger discharge is not dominant. In Sec.~\ref{sec:WaveEqs} we present the separated wave equations for scalars and fermions with non-zero mass and charge. In Sec.~\ref{sec:TransmissionFacs} we use some approximation methods to determine analytic expressions for transmission factors for massive charged scalars and massless charged fermions. To conclude, in Sec.~\ref{sec:Results} we plot our new results for various combinations of parameters and in Sec.~\ref{sec:Conclusions} we discuss the main consequences for LHC phenomenological studies.  
  
\section{The background}\label{sec:background}

In this study we are considering a $(4+n)$-dimensional black hole space-time which is asymptotically flat.
In the context of TeV gravity scenarios for the LHC such as large extra dimensions~\cite{Antoniadis:1990ew,Arkani-Hamed:1998rs,Antoniadis:1998ig,Arkani-Hamed:1998nn} or warped extra dimensions~\cite{Randall:1999vf}, they are expected to form due to the strong gravitational interaction between colliding partons with a centre of mass energy well above $M_{4+n} \sim 1 \ \mathrm{TeV}$~\cite{Argyres:1998qn,Dimopoulos:2001hw,Giddings:2001bu}. This scale is taken to be the fundamental Planck mass in flat $(4+n)$-dimensional space-time with four infinite dimensions and $n$ compactified extra dimensions of typical size orders of magnitude larger than $1 \ \mathrm{TeV}^{-1}$. The four-dimensional Planck mass $M_4$ is an effective quantity at large distances. Its large value arises from integrating out the extra dimensions, giving a large volume factor which scales up $M_{4+n}$ (see for example~\cite{Arkani-Hamed:1998rs}). 

Since all scales involved in the process are assumed to be of the order of $1 \ \mathrm{TeV}$ it follows that the black hole so formed can effectively be treated as a $(4+n)$-dimensional object. In general it will be characterized by a mass $M$, angular momentum $J$ and some Standard Model charges inherited from the colliding partons. In particular, for proton-proton collisions, since quarks are electrically charged it can have a charge $Q$ (in this paper we will not consider colour charges). Furthermore, at formation, it will have higher multipoles associated with asymmetries during the collision. Due to the no hair theorems of general relativity and some estimates in the literature~\cite{Giddings:2001bu}, it is believed that such asymmetries are lost quickly. Thereafter the set $\{M,J,Q\}$ should provide a good description of the black hole. 

Early attempts to model black hole production~\cite{Eardley:2002re,Kohlprath:2002yh,Yoshino:2002br,Yoshino:2002tx,Yoshino:2005hi} and evaporation~\cite{Ida:2002ez,Harris:2003eg,Harris:2005jx,Ida:2005ax,Duffy:2005ns,Casals:2005sa,Cardoso:2005vb,Cardoso:2005mh,Ida:2006tf,Casals:2006xp,Casals:2008pq} focused on mass ($M$) and angular momentum ($J$). Here we are interested in adding charge and particle mass corrections to the evaporation. We start with the projected Myers-Perry metric with mass $M$ and one angular momentum $J$ on the brane, and solve for the Maxwell field keeping the background fixed. It is important to emphasize that we are not interested in a Maxwell field propagating in the bulk of the $(4+n)$-dimensional space-time, but rather a Maxwell field confined to a 4-dimensional brane where all the SM fields propagate\footnote{For a discussion of why this is so, see for example section~3 of \cite{Frost:2009cf}.}.

The background gravitational field is given by the Myers-Perry metric~\cite{Myers:1986un}
\begin{multline}\label{MP_metric}
ds^2=\left(1-\dfrac{\mu}{\Sigma r^{n-1}}\right) dt^2+\dfrac{2a \mu \sin^2{\theta}}{\Sigma r^{n-1}}dt d\phi-\dfrac{\Sigma}{\Delta}dr^2-\\ -\Sigma d\theta^2-\left(r^2+a^2+\dfrac{a^2 \mu \sin^2{\theta}}{\Sigma r^{n-1}}\right)\sin^2{\theta}d\phi^2
-r^2\cos^2\theta d\Omega_n^2 \ , \end{multline}
where
\begin{equation}
\Delta=r^2+a^2-\dfrac{\mu}{r^{n-1}}\,, \hspace{1cm} \Sigma=r^2+a^2\cos^2{\theta} \ ,
\end{equation}
$t$ is a time coordinate, $d\Omega_n^2$ is the metric on an $n$-sphere and $\left\{r,\theta,\phi\right\}$ are spatial spheroidal coordinates. The mass parameter $\mu$ and the oblateness parameter $a$ are related to the physical mass and angular momentum respectively through
\begin{eqnarray}
\dfrac{M}{M_{4+n}}&=&\dfrac{(n+2)}{2}S_{2+n}(2\pi)^{-\frac{n(n+1)}{n+2}}\,M_{4+n}^{n+1}\mu \ , \\
J&=&S_{2+n}(2\pi)^{-\frac{n(n+1)}{n+2}}\,M_{4+n}^{n+2}\,a\,\mu =\dfrac{2}{n+2}\,M a \ .
\end{eqnarray}
where $S_{2+n}$ is the surface area of a $(2+n)$-sphere and we have adopted the PDG convention for the extra-dimensional Planck mass $M_{4+n}$ as in~\cite{Frost:2009cf}. 

By fixing the coordinates $\Omega_n$ we obtain the brane projected metric. This will suffice as an effective metric to describe the gravitational field felt by brane fields. 

\subsection{Determination of the Maxwell field}
\label{sec:MaxwellField}
As a starting approximation, assume the Maxwell field is a perturbation on top of the background gravitational field. We want to solve Maxwell's equations for the vector potential $A_a$ using the metric \eqref{MP_metric}. The combined gravitational plus electromagnetic background can then be coupled to other fields to study the Hawking effect. 

We want the solution to retain the symmetries of the effective four dimensional background. The latter has exactly the same symmetries as the Kerr-Newman solution so we use the same type of ansatz (see for example chapter 3.6 of~\cite{Frolov:1998wf}) 
\begin{equation}\label{KN_ansatz}
A_a dx^a = -Q \dfrac{r}{\Sigma}\left(dt-a\sin^2\theta d\phi\right) \ .
\end{equation} 
where $A_a$ is the vector potential. It can be check that \eqref{KN_ansatz} solves the sourceless Maxwell equations on the brane
\begin{equation}\label{Maxwell1}
D_a F^{ab}=\dfrac{1}{\sqrt{g}}\partial_a\left(\sqrt{g}F^{ab}\right)=0 \ .
\end{equation}
where $F_{ab}=\partial_aA_b-\partial_bA_a$ is the field strength. This result follows since $\sqrt{g}=\Sigma\sin\theta$ is exactly the same as for the Kerr-Newman metric\footnote{Note that $g_{ab}$ is the brane projected metric. Throughout, Latin indices denote curved space-time components whereas Greek indices denote Minkowski space components.}. In addition the identities \[D_{\left[a\right.} F_{\left.bc\right]}=0 \; ,\] where the brackets denote cyclic permutation of indices, are also satisfied. Note how the modified $r^{1-n}$ term in $\Delta$ which gives a $1/r^{2+n}$ gravitational force law away from the black hole, does not affect the stationary brane Maxwell field. This means that a brane charged particle propagating outside the black hole, feels an electric force that scales like $1/r^2$ and a gravitational force that scales like $1/r^{2+n}$.

Gauss' theorem applied to Eq.~\eqref{Maxwell1} allows us to match $Q$ to the physical charge of the black hole
\begin{equation}
\int D_a F^{ab}d\Sigma_b= 4\pi\int d\Sigma_c\sqrt{g} J^c \Rightarrow Q= \int d^3x\sqrt{g} J^0 \ .
\end{equation}
Here we have integrated over spatial hypersurfaces of constant $t$ with normal hypervolume $d\Sigma_b=d^3x\delta_b^0$ and after applying Gauss' theorem integrated the left hand side on a sphere at $r\rightarrow +\infty$.

\subsection{Comments on backreaction}\label{sec:backreact}
From the equivalence principle,  we know that the Maxwell field should also source the right hand side of Einstein's equations through its energy-momentum tensor. In other words the Maxwell field also gravitates and will generate a correction to the metric. Now that we have found a consistent solution of Maxwell's equations on the background we may try to find a self-consistent correction to the gravitational field. Ideally we would have to solve the coupled Einstein-Maxwell equations in the full $(4+n)$-dimensional space with the Maxwell field confined to the brane. This would involve finding a specific mechanism to confine the field. 

In Sect.~\ref{sec:MaxwellField} we started with a simplified framework where the brane is a small perturbation and its geometry is correctly described by the projected Myers-Perry metric~\eqref{MP_metric}. It can be checked by direct computation that the Einstein equations on the brane for that background projected metric are not vacuum like. This is not surprising since the actual vacuum black hole solution lives in $(4+n)$-dimensions. The non-zero components are 
\begin{eqnarray}
{G^{(0)}}_r^r&=& \dfrac{n\mu r^{1-n}}{\Sigma^2} \nonumber \\
{G^{(0)}}_\theta^\theta&=& -\dfrac{{G^{(0)}}_r^r}{2r^2}\left[(n+1)r^2+(n-1)a^2\cos^2\theta\right] \nonumber \\
{G^{(0)}}_\phi^\phi&=& -\dfrac{{G^{(0)}}_r^r}{2r^2\Sigma}\left[(n+1)r^4+(n+3)r^2a^2+(n-3)r^2a^2\cos^2\theta+(n-1)a^4\cos^2\theta\right] \nonumber \\
{G^{(0)}}_t^t&=& \dfrac{{G^{(0)}}_r^r}{2r^2\Sigma}\left[2r^4+(n+3)r^2a^2-(n+1)r^2a^2\cos^2\theta+(n-1)a^4\cos^2\theta\sin^2\theta\right] \nonumber \\
{G^{(0)}}_t^\phi&=& \dfrac{a{G^{(0)}}_r^r}{2r^2\Sigma}\left[(n+1)r^2+(n-1)a^2\cos^2\theta\right] \nonumber \\
{G^{(0)}}_{\phi}^t&=& -{G^{(0)}}_t^\phi\Sigma_0\sin^2\theta \; , 
\end{eqnarray}
where $\Sigma_0=r^2+a^2$. So from the brane point of view, an observer performing gravitational measurements sees a black hole space-time together with an effective fluid due to the embedding into the extra dimensions. 

Before trying to find the corrections to the metric it is useful to note some properties. We expect such a corrected metric to reduce to the projected metric~\eqref{MP_metric} in the $Q=0$ limit and to the Kerr-Newman solution when $n=0$. Furthermore, it should exhibit the same symmetries as the Kerr-Newman metric if we want the Maxwell field to be of the same form as in Eq.~\eqref{KN_ansatz}. 

Compared to the Kerr ($Q=0$) limit, the Kerr-Newman metric is modified by a shift of the mass term $\mu r$ in $\Delta$ to  $\mu r -Q^2$. The term $\mu r$ is related to the gravitational potential which in the chargeless  $(4+n)$-dimensional case is simply replaced by $\mu r^{1-n}$. Similarly, we adopt an ansatz where $\mu r^{1-n}$ is shifted to $\mu r^{1-n}-Q^2$ (or equivalently $\Delta\rightarrow \Delta+Q^2$). This substitution has been noted in a Randall-Sundrum context~\cite{Aliev:2005bi} where $Q^2$ is interpreted as a tidal charge.  Then the effective brane metric ansatz is 
\begin{multline}\label{ansatz_metric}
ds_{(4)}^2=\left(1-\dfrac{\mu r^{1-n}-Q^2}{\Sigma}\right) dt^2+\dfrac{2a (\mu r^{1-n}-Q^2) \sin^2{\theta}}{\Sigma}dt d\phi-\dfrac{\Sigma}{\Delta}dr^2-\\ -\Sigma d\theta^2-\left(r^2+a^2+\dfrac{a^2 (\mu r^{1-n}-Q^2) \sin^2{\theta}}{\Sigma }\right)\sin^2{\theta}d\phi^2 \ , \end{multline} 
and
\[\Delta=r^2+a^2-\mu r^{1-n}+Q^2 \ .\]
Remarkably, explicit evaluation of the Einstein tensor for this metric yields
\begin{equation}\label{eq:braneEinstein}
{G}_a^b={G^{(0)}}_a^b+8\pi T_a^b \; ,
\end{equation}
where 
\begin{equation}
8\pi T_a^b=\left(\begin{array}{cccc} -\dfrac{Q^2(\Sigma_0+a^2\sin^2\theta)}{\Sigma^2}& 0&0&-\dfrac{2aQ^2}{\Sigma^3} \\ 0&-\dfrac{Q^2}{\Sigma^2}&0&0 \\ 0&0&\dfrac{Q^2}{\Sigma^2}&0 \\ \dfrac{2aQ^2\Sigma_0 a^2\sin^2\theta}{\Sigma^3}&0&0&\dfrac{Q^2(\Sigma_0+a^2\sin^2\theta)}{\Sigma^3}\end{array}\right)
\end{equation} is the energy momentum tensor for the Maxwell field obtained in the previous section, as computed from the definition
\begin{equation}
T_a^b=\dfrac{1}{4\pi}\left(F_{ac} F^{bc}-\dfrac{1}{4}\delta_a^b F_{cd} F^{cd}\right) \ .
\end{equation}
This shows how the brane metric ansatz we have chosen reproduces exactly the gravitational field generated by the Maxwell field while keeping the extra contribution from the embedding into the bulk untouched. This indicates that we can consistently add the Maxwell field on the brane and correct the brane metric accordingly.

Even though the effective metric~\eqref{ansatz_metric} can't be the full solution we can regard it as a first approximation which is physically consistent (for a rigorous study in the second Randall-Sundrum model see~\cite{Chamblin:2000ra,Aliev:2005bi}). To solve the problem of the backreaction exactly, we would have to construct a bulk energy momentum tensor for the Maxwell field, with some typical thickness, and solve the bulk Einstein equations. This would give the effect of the four dimensional brane Maxwell field on the bulk geometry as well as the brane. Keeping in mind the ansatz above it is tempting to assume that the physical metric will have the form
\begin{multline}
ds_{(4)}^2=\left(1-\dfrac{\mu r^{1-n}-Q^2(\Omega_n)}{\Sigma}\right) dt^2+\dfrac{2a (\mu r^{1-n}-Q^2(\Omega_n)) \sin^2{\theta}}{\Sigma}dt d\phi-\dfrac{\Sigma}{\Delta}dr^2-\\ -\Sigma d\theta^2-\left(r^2+a^2+\dfrac{a^2 (\mu r^{1-n}-Q^2(\Omega_n)) \sin^2{\theta}}{\Sigma }\right)\sin^2{\theta}d\phi^2 +r^2\cos^2\theta d\Omega_n^2 \ , \end{multline} 
where now $Q^2$ is a function of the transverse bulk coordinates $\Omega_n$ such that
\[Q(\Omega_n)=\left\{\begin{array}{ll} Q & \ \ \  \mathrm{if \ } \Omega_n \mathrm{\ on\  the\  brane}\\ 0 & \ \ \ \mathrm{otherwise} \end{array}\right. \ .\]
If we imagine a brane with thickness $\epsilon$ such that the charge function $Q^2(\Omega_n)$ drops suddenly where the brane ends, then this choice ensures the vacuum Einstein equations are obeyed in the bulk, as well as on the brane (together with the Maxwell field). The only addition is a sharp  $\delta$ function like energy momentum tensor where the brane ends. This can be checked explicitly in the 5-dimensional case by using a generic function $Q(\chi)$ ($\chi$ is the fifth dimensional coordinate) and applying the Gauss-Codazzi equations to obtain brane Einstein equations at each hyperslice parallel to the $\chi=0$ brane. If the profile chosen is flat inside the brane ($Q(\chi)=Q$) and drops suddenly to zero at some $\chi_\epsilon$, then we obtain terms which are proportional to derivatives of $Q(\chi)$ at $\chi_\epsilon$. These extra contributions at $\chi_\epsilon$ spoil the construction but nevertheless we can ignore them or assume they are somehow related to the mechanism that keeps the fields confined to the $4$-dimensional brane. 

Regardless of these problems, note that the charge introduced in~\eqref{ansatz_metric} consistently reduces the size of the black hole event horizon on the brane as we would expect for a charged black hole. Furthermore the Maxwell field, which is independent of $\Delta$, produces terms in the geodesics which reproduce exactly the usual 4-dimensional electric force. This is certainly a feature we want to keep. Finally, for LHC black holes we will see that the $Q^2$ in the metric is actually a small perturbation. So the charge shouldn't disturb the bulk geometry much and to first order this effective brane metric should be a good approximation. 

\section{Systems of units and orders of magnitude}
\label{sec:OrdersOfMagnitude}
In this section we find the relation between the black hole parameters and the corresponding physical quantities in terms of well known constants, as well as the coupling of charged test particles. Note that for simplicity, we have been working in a natural system of units where all dimensionful quantities come in fact divided by the appropriate Planck unit factor. For example lengths come divided by $M_{4+n}^{-1}$ and masses by $M_{4+n}$. Similarly any field comes divided by the appropriate ``Planck quantity''. Then the charge $Q$ becomes a dimensionless quantity describing the strength of the electric field with respect to some reference charge. The precise value of this parameter is found by matching to a known limit. Anticipating the result we write $Q=Z\sqrt{\alpha}$ where $\sqrt{\alpha}$ is the fundamental charge and $Z$ is the charge of the black hole in units of $\sqrt{\alpha}$. For the purpose of matching $Q$, the rotation parameter can be set to zero.

Let's start by looking at geodesics for charged particles. They are obtained from the action principle
\begin{equation}
S=\int d\lambda\left(\dfrac{1}{2}\dfrac{dx^{a}}{d\lambda}\dfrac{dx_{a}}{d\lambda}+q\dfrac{dx^{a}}{d\lambda}A_{a}\right)
\end{equation}
where $q=z\sqrt{\alpha}$ is the charge of the test particle. The coupling $\sqrt{\alpha}$ can be found by taking the non-relativistic limit. If we define
the generalised momentum
\begin{equation}\label{Pcanonical}
P_{a}=\dfrac{d\mathcal{L}}{d\dot{x}^{a}}=\dfrac{dx_{a}}{d\lambda}+qA_{a} \; ,
\end{equation}
conservation of the Hamiltonian $\mathcal{H}\equiv\mathcal{L}-P_{a}\dot{x}^{a}$ reduces to the 4-momentum constraint
\begin{equation}
p_{a}p^{a}=m^{2} \; ,
\end{equation}
where $p^{a}=dx^{a}/d\lambda$. The geodesic
equation coupled to electromagnetism is obtained by variation of the
action:
\begin{eqnarray}
\dfrac{d^{2}x^{a}}{d\lambda^{2}}+\Gamma_{bc}^{a}\dfrac{dx^{b}}{d\lambda}\dfrac{dx^{c}}{d\lambda}+qF_{b}^{a}\dfrac{dx^{b}}{d\lambda} & = & 0 \; ,
\end{eqnarray}
 where $\Gamma^a_{bc}$ are the Christoffel symbols. We consider radial
geodesics $d\theta/d\lambda=d\phi/d\lambda=0$. In four dimensions, the non-trivial equations are
\begin{eqnarray}
\dfrac{d^{2}r}{d\lambda^{2}}+\dfrac{1}{2}m^{2}U^{\prime}-\alpha U\dfrac{zZ}{r^{2}}E & = & 0\\
\dfrac{d^{2}t}{d\lambda^{2}}+\left(U^{\prime}E+\alpha\dfrac{zZ}{r^{2}}\right)\dfrac{1}{U}\dfrac{dr}{d\lambda} & = & 0 \; ,
\end{eqnarray}
 with $U=\Delta/r^2$. In the non-relativistic limit $dt/d\lambda=E\sim m$, therefore
$dt\sim m\, d\lambda$ and 
\begin{eqnarray}
m\dfrac{d^{2}r}{dt^{2}} & = & -\dfrac{mM}{r^{2}}+\alpha\dfrac{zZ}{r^{2}}+\alpha m\dfrac{Z^{2}}{r^{3}}+O(r^{-4}) \; ,
\end{eqnarray}
 giving respectively the Newtonian and Coulomb force laws and the first relativistic correction due to the gravitational effect of the Maxwell field. To match $\alpha$ in four dimensions, put back all length scales in terms of Planck units explicitly (note that $l_{4}$, and $l_{4+n}$ are Planck lengths)
\begin{eqnarray}\label{eq:match_geodesic1}
m\dfrac{d^{2}r}{dt^{2}} & \dfrac{1}{M_{4}^{2}}=\dfrac{m}{M_{4}} & \left(-\dfrac{M}{M_{4}}\dfrac{l_{4}^{2}}{r^{2}}+\alpha\dfrac{M_{4}}{m}zZ\dfrac{l_{4}^{2}}{r^{2}}+\alpha Z^{2}\dfrac{l_{4}^{3}}{r^{3}}+\ldots\right) \; ,
\end{eqnarray}
 where the extra relativistic correction is suppressed
by one more power of $l_{4}/r$. Rewriting
the previous equation and setting the masses to electron masses and charges $z,Z=1$ (i.e. the unit is the electron charge) 
\begin{eqnarray}
m\dfrac{d^{2}r}{dt^{2}} & \dfrac{1}{M_{4}^{2}}= & -\dfrac{m_{e}}{M_{4}}\left(\dfrac{m_{e}}{M_{4}}-\alpha\dfrac{M_{4}}{m_{e}}\right)\dfrac{l_{4}^{2}}{r^{2}}+\ldots \; . \end{eqnarray}
The ratio of electric to gravitational force between electrons gives
\begin{equation}\label{eq:matching4D} 
\alpha=\dfrac{F_{e}}{F_{g}}\left(\dfrac{m_{e}}{M_{4}}\right)^{2}=\dfrac{\dfrac{e^{2}}{4\pi\epsilon_{0}}}{Gm_{e}^{2}}\left(\dfrac{m_{e}}{M_{4}}\right)^{2}=\dfrac{e^{2}}{4\pi\epsilon_{0}}\simeq\dfrac{1}{137}
\end{equation}
 as expected. Eq.~\eqref{eq:matching4D} emphasizes how the electric force $F_e=\alpha F_g(M_{4}/m_e)^2$ in 4-dimensions is orders of magnitude stronger than the gravitational force. This is simply a statement of the hierarchy problem mentioned in the introduction. However the same cannot apply in $\mathrm{TeV}$ gravity scenarios where all forces are controlled by the same scale, so gravity becomes stronger at short distances. Thus it is crucial to determine the relative strength of the $(4+n)$-dimensional gravitational force and the electric force.

Now let us rewrite Eq.~\eqref{eq:match_geodesic1} using $M_{4+n}$
\begin{multline}
m\dfrac{d^{2}r}{dt^{2}} \dfrac{1}{M_{4+n}^{2}}= \\ \dfrac{m}{M_{4+n}} \left[-\left(\dfrac{M_{4+n}}{M_{4}}\right)^2\dfrac{M}{M_{4+n}}\dfrac{l_{4+n}^{2}}{r^{2}}+\alpha\dfrac{M_{4+n}}{m}zZ\dfrac{l_{4+n}^{2}}{r^{2}}+\left(\dfrac{M_{4+n}}{M_{4}}\right)^2\alpha Z^{2}\dfrac{l_{4+n}^{3}}{r^{3}}+\ldots\right]
\end{multline}
The first and third contributions, which are due to the gravitational fields of the mass $M$ and the charge $Q$, are suppressed by the same power of $M_{4+n}/M_{4}$. However, as we approach short distances, the gravitational coupling must become higher dimensional,\footnote{This can be checked explicitly by using the brane metric~\eqref{ansatz_metric}.} gravity becomes strong and the suppression factors will disappear. Note however, that since the Maxwell field is confined to the brane, the $r$-power in the third term (which is associated with the gravitational effect of the charge) remains the same. As for the second term, it is associated with the electric force between the test particle and the charged body so it must remain the same, again because the Maxwell field is confined to the brane and the magnitude of the electric force cannot change at shorter distances.  

This qualitative discussion agrees with the short distance geodesic equation obtained from Eq.~\eqref{ansatz_metric}
\begin{multline}
m\dfrac{d^{2}r}{dt^{2}} \dfrac{1}{M_{4+n}^{2}}= \dfrac{m}{M_{4+n}} \left[-(n+1)\mu M_{4+n}^{n+1} \dfrac{l_{4+n}^{n+2}}{r^{n+2}}+\alpha\dfrac{M_{4+n}}{m}zZ\dfrac{l_{4+n}^{2}}{r^{2}}+\alpha Z^{2}\dfrac{l_{4+n}^{3}}{r^{3}}+\ldots\right] \ .
\end{multline}
The first term is correctly modified to a higher dimensional force law, the second term remains the same and the third term is controlled by the same power or $r$ but without the suppression factor $\left(M_{4+n}/M_{4}\right)^2$.

It is worth noting that for LHC black holes, which can be produced with a maximum charge of $|Z|=4/3$, the fine structure constant factor of $~1/137$ makes the $Q^2$ contribution to the metric small (unless the BH happens to charge up to $|Z|\sim 10$ during the evaporation).

\section{Evaporation -- coupling the Maxwell field} \label{sec:Evaporation}
Now that we have constructed a physically reasonable background we can move on to discuss the effects of charge on the evaporation. The coupling of the Einstein-Maxwell background is straightforward through covariantisation. The action principle for the scalar field is
\begin{equation}\label{eq:scalar_action}
S_\Phi=\int \mathrm{d}^4x\sqrt{g}\left(\dfrac{1}{2}D_a \Phi D^b \Phi -\dfrac{1}{2}\mu^2\Phi^2\right) \; ,
\end{equation}
with
\begin{equation}
D_a=\nabla_a+iqA_a \; ,
\end{equation}
where $\nabla_a$ is the space-time covariant derivative and we have anticipated the matching of the coupling $q$ to the one introduced before. Note that from now on we use $\mu$ for the mass of the field\footnote{This is to avoid confusion with the azimuthal quantum number $m$ introduced later on.} and will eliminate the explicit dependence on the $\mu$ parameter of Eq.~\eqref{ansatz_metric} by changing units -- see Eq.~\eqref{eq:new_units} below. Variation of \eqref{eq:scalar_action} gives the wave equation
\begin{equation}\label{eq:scalar_wave}
\left(D^b D_b+\mu^2\right)\Phi=0 \ .
\end{equation}

To check the coupling is correct we take the classical limit. In flat space-time consider a slowly varying vector potential $A_a$, set 
\begin{equation} 
\Phi\sim e^{iS}
\end{equation} 
and identify the mechanical 4-momentum of the classical particle with $p_a=-\nabla_a S-qA_a$. Then to leading order, Eq.~\eqref{eq:scalar_wave} gives the mass-shell condition \[ p_a p^a = \mu^2\ .\] Conversely, $P_a=-\nabla_a S$ is the usual canonical momentum of the classical particle so we match $q$ to $z\sqrt{\alpha}$ as in Eq.~\eqref{Pcanonical}. This coupling agrees with well studied cases (see for example~\cite{Gibbons:1975kk,Page:1977um,Page:1976jj,Nakamura:1976nc}).

For fields with higher spin the procedure is exactly the same. For a Dirac field we write the action
\begin{equation}
S_{\Psi}=\int \mathrm{d}^4x\sqrt{g}\bar{\Psi}\left(\slash{D}-\mu\right)\Psi \; ,
\end{equation}
where
\begin{equation}
\slash{D}=\gamma^a\left(\nabla_a+iqA_a\right) \; ,
\end{equation}
with
\begin{equation}
\gamma^a\gamma^b+\gamma^b\gamma^a=2g^{ab} \; .
\end{equation}
$\bar{\Psi}=\Psi^{\dagger}\gamma^0$
and the spinor covariant derivative is
\begin{equation}
\nabla_a=\partial_a+\dfrac{1}{8}\omega_{\mu\nu a}\left[\gamma^\mu,\gamma^\nu\right] \; ,
\end{equation}
where the gamma matrices $\gamma^\mu$ are in flat space-time and $\omega_{\mu\nu a}$ is the spin connection.

\subsection{Hawking radiation}\label{sec:HawkingRad}
Since the pioneering work of Hawking~\cite{Hawking:1974sw} several studies in the literature have examined the quantisation of various fields in black hole backgrounds which are analytically similar to the one we are using~\cite{Unruh:1974bw,Gibbons:1975kk,Candelas:1981zv,Ottewill:2000qh,Casals:2005kr}. In particular the metric~\eqref{ansatz_metric} and the Maxwell field~\eqref{KN_ansatz} are similar in form to Kerr-Newman, so the quantisation procedure is formally the same and we will not repeat it referring the interest reader to~\cite{Unruh:1974bw,Gibbons:1975kk,Candelas:1981zv,Ottewill:2000qh,Casals:2005kr}. 

Before presenting a summary of the physical quantities which are relevant to our study, it is convenient to adopt horizon radius units where $r_H=1$ ($r_H$ is defined as the largest positive root of $\Delta=0$). The mapping of parameters is\footnote{Note that $\mu$ here is the mass of the particle. The $\mu$ in $\Delta$ has been eliminated through this change of system of units.}
\begin{equation} \label{eq:new_units}
\begin{array}{ccc}
\dfrac{r}{r_H}\rightarrow r &\hspace{1cm} \dfrac{a}{r_H}\rightarrow a \\ \vspace{-2mm}
& \\
\omega r_H \rightarrow \omega &\hspace{1cm}  \mu r_H \rightarrow \mu \\ \vspace{-2mm}
& \\
q r_H \rightarrow q &\hspace{1cm}  \dfrac{Q}{r_H}\rightarrow Q 
\end{array}
\end{equation} 
so $\Delta$ becomes
\begin{equation}
\Delta = r^2+a^2+Q^2-(1+a^2+Q^2)r^{1-n} \ .
\end{equation}
The main result of Hawking's original paper is that black holes emit a continuous flux of particles. In our system of units the fluxes of particle number, energy, angular momentum and charge are respectively~\cite{Hawking:1974sw}
\begin{eqnarray}
\frac{d^2 N_q}{dt d\omega} &=& \frac{1}{2\pi} \sum_{j=|s|}^\infty \sum_{m = -j}^{j} \frac{1}{\exp(\tilde{\omega}/T_H) \pm 1} \mathbb{T}^{(4+n)}_{k}(\omega,\mu, a,q,Q)\;,  \label{eq-flux-spectrum}\\
\frac{d^2 E_q}{dt d\omega} &=& \frac{1}{2\pi} \sum_{j=|s|}^\infty \sum_{m = -j}^{j} \frac{\omega}{\exp(\tilde{\omega}/T_H) \pm 1} \mathbb{T}^{(4+n)}_{k}(\omega,\mu, a,q,Q)\;,  \label{eq-power-spectrum}\\
\frac{d^2 J_q}{dt d\omega} &=& \frac{1}{2\pi} \sum_{j=|s|}^\infty \sum_{m = -j}^{j} \frac{m}{\exp(\tilde{\omega}/T_H) \pm 1} \mathbb{T}^{(4+n)}_{k}(\omega,\mu, a,q,Q)\;,\label{eq-angular-spectrum}\\
\frac{d^2 Q_q}{dt d\omega} &=& \frac{1}{2\pi} \sum_{j=|s|}^\infty \sum_{m = -j}^{j} \frac{q}{\exp(\tilde{\omega}/T_H) \pm 1} \mathbb{T}^{(4+n)}_{k}(\omega,\mu, a,q,Q)\;,\label{eq-charge-spectrum}
\end{eqnarray}
where $\tilde{\omega}=\omega-m\Omega_H-q \Phi_H$, $k=\{j,m\}$ are the angular momentum quantum numbers, $s$ is the helicity of the particle 
\begin{equation}
T_H=\dfrac{(n+1)+(n-1)(a^2+Q^2)}{4\pi(1+a^2)r_H} \; ,
\end{equation}
and the signs $\pm$ are for fermions and bosons respectively.  $\Omega_H=a/(1+a^2)$ and $\Phi_H=Q/(1+a^2)$ are the angular velocity and electric potential of the horizon respectively. $\Phi_H$ can be defined using the timelike Killing vector at the horizon. For metric~\eqref{ansatz_metric}, we can pick a Killing vector field which is timelike at a given point, using the two Killing vector fields $\mathbf{k}_t=e_t$ and $\mathbf{k}_\phi=e_\phi$. We denote such a vector 
\begin{equation}
\mathbf{k}_p=e_t+\Omega_{p}e_\phi \; ,
\end{equation}
where the subscript $p$ labels a space-time point. Then if $\mathbf{k}_p$ is timelike at $p$, $\Omega_{-}<\Omega_p<\Omega_{+}$ where
\begin{equation}
\Omega_{\pm}=\dfrac{-g_{t\phi}\pm\sqrt{g_{t\phi}^2-g_{tt}g_{\phi\phi}}}{g_{\phi\phi}} \; .
\end{equation}
At the horizon $\Omega_{+}=\Omega_{-}=\Omega_H$, so in some sense there is a natural vector field which defines the timelike direction close to the horizon. The electric potential at the horizon $\Phi_H$ is defined as the projection of the $A_{a}$ field along $\mathbf{k}_H^a$. It can also be shown that $\Omega_H$ corresponds to the angular velocity of a physical observer close to the horizon whose frame is dragged by the gravitational field of the rotating black hole~\cite{Bardeen:1972fi}.  

Finally $\mathbb{T}^{(4+n)}_k$ are the so called transmission factors defined as the fraction of an incident wave from infinity which is absorbed by the black hole. The boundary conditions are such that close to the horizon the wave is purely ingoing for the above physical observers~\cite{Bardeen:1972fi}.

Before going into the details of calculating fluxes, it is instructive to look at an estimate from the Schwinger formula for fermions~\cite{Schwinger:1951nm}:
\begin{equation}\label{schwinger_estimate}
\dfrac{dN}{dVdt}=\dfrac{q^2E^2}{\pi^2}\sum_{n=1}^{+\infty}{\dfrac{e^{-\frac{n\pi\mu^2}{qE}}}{n^2}}\simeq \dfrac{q^2E^2}{6} \; ,
\end{equation}
where we took the small mass limit and $E$ is the electric field. Eq.~\eqref{schwinger_estimate} is valid in flat space for a uniform electric field, and it gives the rate of production of opposite charge pairs due to the electric field only. For~\eqref{KN_ansatz}, we know that the electric field drops like $1/r^2$ so strictly speaking this formula is not valid. Nevertheless we can still use~\eqref{schwinger_estimate} to estimate the contribution of the background electric field to particle production and compare it with the contribution from the gravitational field alone (i.e. the typical Hawking flux for a neutral black hole). A rough estimate is obtained by considering the electric field at the horizon and a volume of order $(2r_H)^3$ around the black hole. Using our system of units and noting that the electric field at the horizon is $E_H\sim Q/r_H^2$ we get
\begin{equation}
\dfrac{dN}{dt} r_H \sim q^2Q^2 = z^2Z^2\alpha^2\simeq z^2Z^2 10^{-5} \ .
\end{equation}
So for order $\sim 1$ charges we get a very small rate when compared to the typical Hawking fluxes for a neutral black hole (which are of order $\sim 1$). This indicates that pair production due to the gravitational field is much stronger than pair production due to the electric field. So the common claim that TeV-scale black holes lose their charges earlier in their lifetime, is not necessarily true on the basis of Schwinger discharge alone. Below we confirm this result with a more detailed calculation.    

\subsection{Wave equations}\label{sec:WaveEqs}
In this section we present the separated scalar and fermion wave equations. Then we solve them with appropriate boundary conditions to obtain transmission coefficients, using an approximate method which is valid for low energies~\cite{Creek:2007tw,Creek:2007sy,Creek:2007pw,Creek:2006ia} and a WKB approximation in the high energy limit. Our new results for charged and massive fields reduce in some limits, to the ones in~\cite{Creek:2007tw,Creek:2007sy,Creek:2007pw,Creek:2006ia,Gibbons:1975kk} which can be used as a check.

\subsubsection{The scalar field}
Using Eq.~\eqref{ansatz_metric} and the separation ansatz $\Phi=e^{-i\omega t+im\phi}R(r)S(\theta)$ we obtain the radial equation
\begin{equation}\label{eq:radial2s0}
\Delta\dfrac{d}{dr}\left(\Delta \dfrac{dR}{dr}\right)+\left(K^2-\Delta U\right)R=0 \; ,
\end{equation} 
where
\begin{eqnarray}
K&=& \omega\Sigma_0-a m-qQr \\
U&=& \mu^2r^2+\Lambda_{c,j,m}+\omega^2a^2-2a\omega m \; .
\end{eqnarray}
The angular equation has the same form as in the chargeless case
\begin{equation}\label{eq:angular2s0}
\dfrac{1}{\sin\theta}\dfrac{d}{d\theta}\left(\sin\theta\dfrac{dS}{d\theta}\right)+\left(c^2\cos^2\theta-\dfrac{m^2}{\sin^2\theta}+\Lambda_{c,j,m}\right)S=0 \; ,
\end{equation}
where $c^2=a^2(\omega^2-\mu^2)$ and $\Lambda_{c,j,m}$ is the angular eigenvalue. In particular when evaluating our analytic results for transmission factors, we will use well known expansions for the angular eigenvalue~\cite{Berti:2005gp}.

Equation~\eqref{eq:radial2s0}, is similar to the chargeless case with the additional terms:
\begin{enumerate}
\item $Q^2$ in $\Delta$, which changes the location of the horizon and therefore the Hawking temperature of the black hole.
\item $qQr$ in $K$ which is related to the electric potential. 
\end{enumerate}
In $K$, $\omega$ is shifted  by 
\begin{equation}-\dfrac{am}{r^2+a^2}-\dfrac{qQr}{r^2+a^2} \ . \end{equation}
Evaluated at the horizon, both quantities are associated with the well known phenomenon of superradiance~\cite{Bardeen:1972fi,Teukolsky:1973ha,Press:1973zz,Starobinskii:1973,StarobinskiiChurilov:1973}, i.e. for $\tilde{\omega}<0$ an incident wave from infinity will be scattered back with a larger amplitude. This factor is also present in the expressions for the fluxes such as~\eqref{eq-flux-spectrum} where the Boltzmann suppression factor in the denominator becomes smaller for supperradiant modes.

\subsubsection{The Dirac field}
 For a fermion field the standard procedure to separate the wave equation is to use the Newman-Penrose formalism. The method has been developed for the Kerr metric by Chandrasekhar~\cite{Chandrasekhar:1976ap} and applied to the Kerr-Newman background by Page~\cite{Page:1976jj}. Page points out how a simple substitution of some of Chandrasekhar's quantities suffices to obtain separated equations for the fermion field with charge. Below we state the final result and refer the technical details to references~\cite{Chandrasekhar:1976ap,Page:1976jj,Chandrasekhar:1985kt}.

The separated wave equation for a massive charged Dirac field relies on the ansatz $\Psi =e^{-i(\omega t -m\phi) }\chi(r,\theta)$  where 
\begin{equation}
\chi=\left(\begin{array}{c}(r-ia\cos\theta)^{-1}P_{-1/2}(r)S_{-1/2}(\theta) \vspace{1mm}\\ \sqrt{2}\Delta^{-1/2}P_{+1/2}(r)S_{+1/2}(\theta)\vspace{1mm} \\ \sqrt{2}\Delta^{-1/2}P_{+1/2}(r)S_{-1/2}(\theta)\vspace{1mm} \\-(r+ia\cos\theta)^{-1}P_{-1/2}(r)S_{+1/2}(\theta) \end{array}\right) \; .
\end{equation}
The radial and angular equations are
\begin{equation}\label{eq:2s1radial}
\Delta^{\frac{1}{2}}\left(\dfrac{d}{dr}-2s i\dfrac{K}{\Delta}\right)P_{-s}=\left(\lambda+2is \mu r\right)P_{s}
\end{equation}
and
\begin{equation}\label{eq:2s1angular}
\left[\dfrac{d}{d\theta}+2s\left(a\omega\sin\theta-\dfrac{m}{\sin\theta}\right)+\frac{1}{2}\cot\theta\right]S_{-s}=\left(2s\lambda+a\mu\cos\theta\right)S_{s}
\end{equation}
where $\lambda$ is the angular eigenvalue. To make contact with well know limits, it is useful to obtain second order radial and angular equations by elimination (note that the prime $\prime$ denotes $d/dr$):
\begin{multline}\label{master_radial}
\dfrac{d^2P_{s}}{dr^2}+\left((1-|s|)\dfrac{\Delta^{\prime}}{\Delta}+\dfrac{2is\mu}{\lambda-2is\mu r}\right)\dfrac{dP_{s}}{dr}+\\ +\left[\dfrac{K^2}{\Delta^2}-i s \dfrac{K}{\Delta}\dfrac{\Delta^{\prime}}{\Delta}-\dfrac{4s^2\mu}{\lambda-2i s \mu r}\dfrac{K}{\Delta}+\dfrac{2isK^{\prime}-\lambda^2-\mu^2r^2}{\Delta}\right]P_{s}=0
\end{multline}
and
\begin{multline}\label{master_angular}
\dfrac{1}{\sin\theta}\dfrac{d}{d\theta}\left(\sin\theta\dfrac{dS_{s}}{d\theta}\right)+\dfrac{a\mu\sin\theta}{-2s\lambda+a\mu\cos\theta}\left(\dfrac{d}{d\theta}-2s\left(a\omega\sin\theta-\frac{m}{\sin\theta}\right)+\dfrac{\cot\theta}{2}\right)S_{s}+ \\ +\left[a^2(\omega^2-\mu^2)\cos^2\theta-2sa\omega\cos\theta-\dfrac{\left(m+s\cos\theta\right)^2}{sin^2\theta}+\lambda^2-a^2\omega^2+2a\omega m-|s|\right]S_{s}=0 \; .
\end{multline}
Here $s=\pm 1/2$. In the zero mass limit we recover a radial equation with the same analytic form as in references~\cite{Casals:2006xp,Ida:2006tf} except for the extra term in $K^{\prime}$. Similarly the angular equation is exactly the same as for the spin-half spheroidal functions. Again we can use the expansions in~\cite{Berti:2005gp} for the angular eigenvalues. 

Finally setting the rotation parameter $a$ to zero, note that the angular equation is the same with or without mass and charge. Then the angular eigenvalue takes a closed form and we don't need to integrate the angular equation to study the effects of both mass and charge. This simplification was explored for example in Page's paper in four dimensions~\cite{Page:1977um}. 

\subsection{Transmission factors}\label{sec:TransmissionFacs}
In this section we present analytic approximations for the transmission factors. We are interested in obtaining the main qualitative features of including particle mass and charge using approximations which are valid at low and high energies. In particular the low energy approximation has been shown to give good results even in the intermediate energy regime for spins up to one \cite{Creek:2007tw} so we will obtain a good overall qualitative picture of how the transmission factors behave. Below we present the main steps of the calculation adapted to our problem and refer to details in~\cite{Creek:2007tw}. The method consists of writing down approximations for the radial equation in two regions: one near the horizon (Near Horizon solution) and the other one far from it (Far Field solution). This provides two analytic approximations which hold exactly close to the horizon and far from it respectively. The final step is to extrapolate them into a common intermediate region to be matched. Below we summarize the solutions and keep track of the conditions of validity. A more detailed numerical analysis, which will be useful for improving black hole event generators, is currently in progress~\cite{Sampaio:2009}.

\subsubsection{Near horizon equation}
Equations~\eqref{master_radial} and~\eqref{master_angular} are valid for both spin-zero and spin-half fields. The analytic approximations we use are valid for the massive charged scalar field, but however it turns out they only work for the massless limit of charged fermions. Therefore we work with the radial equation~\eqref{master_radial} but set $\mu=0$ for fermions.  

Following~\cite{Creek:2007tw} close to the horizon define the quantities
\begin{eqnarray}\label{NH_approximations}
f&\equiv &\dfrac{\Delta}{r^2+a^2+Q^2} \nonumber \\
A&\equiv &n+1+(n-1)\dfrac{a^2+Q^2}{r^2}\simeq \left.A\right|_{r=1}\equiv A_* \nonumber\\
B&\equiv &1-|s|+\dfrac{2|s|+n(r^2+a^2+Q^2)}{r^2A}-\dfrac{4(a^2+Q^2)}{r^2A^2}\simeq \left.B\right|_{r=1}\equiv B_* \nonumber \\
P&\equiv & \dfrac{K^2}{r^2A^2}\simeq \dfrac{\omega(1+a^2)-am-qQ}{A_*}\equiv p \nonumber \\
D&\equiv& \dfrac{r^2+a^2+Q^2}{r^2A^2}\left[\lambda^2+\mu^2r^2\delta_{s,0}+2isqQ-4is\omega r\right]\simeq \left.D\right|_{r=1}\equiv D_* \; .
\end{eqnarray}
Then Eq.~\eqref{master_radial} is equivalent to
\begin{equation}\label{eq:near_horizon_radial}
f(1-f)\dfrac{d^2R}{df^2}+(1-Bf)\dfrac{dR}{df}+\left[\dfrac{P^2-isP}{f}+\dfrac{P^2-isP-D}{1-f}\right]R=0 \;
\end{equation}
using the approximations on the right hand side of each line of~\eqref{NH_approximations}. The latter are equivalent to the condition $r-1\ll 1$. Eq.~\eqref{eq:near_horizon_radial} can be solved in terms of hypergeometric functions. The general solution is a combination of two linearly independent hypergeometric functions. According to the general treatment in~\cite{Bardeen:1972fi,Teukolsky:1973ha,Gibbons:1975kk}, at the horizon, the wave must be purely ingoing. This implies $R\sim e^{-ipr_*}$ with $r_*$ the tortoise coordinate defined as $dr_*=dr/f$. Then, it can be shown that the convergent solution with this boundary condition is  
\begin{equation}
R_{NH}= f^{\alpha}(1-f)^{\beta}F(a,b,c;f) \; ,
\end{equation}
where
\begin{eqnarray}
\alpha &=& \dfrac{|s|-s}{2}-ip \nonumber \\
\beta&=& 1-\frac{|s|+B_*}{2}-\sqrt{\left(1-\frac{|s|+B_*}{2}\right)^2-p^2+isp+D_*} \nonumber \\
a&=& \alpha+\beta-1+B_* \nonumber \\
b&=& \alpha+\beta \nonumber \\
c&=& 1-|s|+2\alpha \; .
\end{eqnarray}
In the next section an extrapolation of this solution away from the horizon will be needed, i.e. around $f\rightarrow 1\Rightarrow 1-f\simeq (1+a^2+Q^2)/r^{n+1} $. Note that the larger the value of $n$, the more consistent this condition is with $r-1\ll 1$ so the terms neglected in approximations~\eqref{NH_approximations} become less important\footnote{This improvement of the approximation for large $n$ has been noted in~\cite{Creek:2007tw}.}. Using some identities that relate hypergeometric functions with argument $f$ to argument $1-f$ and expanding around $f=1$ we obtain~\cite{abramowitz} (up to an overall normalisation constant)
\begin{equation}\label{stretch_NH}
R\rightarrow A_1r^{-(n+1)\beta}+A_2r^{-(n+1)(2-\beta-B_*)} \; ,
\end{equation}
with
\begin{eqnarray}
A_1&=&\dfrac{(1+a^2+Q^2)^\beta\Gamma(c-a-b)}{\Gamma(c-a)\Gamma(c-b)} \nonumber \\
A_2&=&\dfrac{(1+a^2+Q^2)^{2-|s|-\beta-B_*}\Gamma(a+b-c)}{\Gamma(a)\Gamma(b)} \; .
\end{eqnarray}
When matching powers of $r$ in the next section we will have to make the approximations $\omega,a,Q,\mu \ll 1$. Then
\begin{eqnarray}\label{eq:powers_match}
-(n+1)\beta & \simeq & -\dfrac{1}{2}+\sqrt{\dfrac{1}{4}+\lambda^2} \nonumber \\
-(n+1)(2-\beta-B_*) & \simeq & -\dfrac{1}{2}-\sqrt{\dfrac{1}{4}+\lambda^2} \ .
\end{eqnarray}
For $a\ll 1$ the neglected terms are of order $\omega^2$ or $\mu^2$. So taking into account the leading behaviour of $\lambda^2$ when $s=|s|$ the approximation is equivalent to 
\begin{equation}\label{eq:condition_ell}
\omega,\mu \ll \sqrt{\ell(\ell+1)+2|s|}\; .
\end{equation} So the larger the $\ell$ and $|s|$ the wider the energy range where the approximations work.

\subsubsection{Far field solution and low energy matching}
Away from the black hole  $r\rightarrow +\infty$ we approximate $\Delta\simeq r^2$ and 
\begin{equation}\label{FF_K}
\dfrac{K}{\Delta}\simeq \omega -\dfrac{qQ}{r}+\dfrac{\omega(1+a^2+Q^2)}{r}\delta_{n,0} \; .
\end{equation} 
Eq.~\eqref{FF_K} contains: the energy, a long range electric potential and a long range gravitational potential in four dimensions. Keeping terms up to order $1/r^2$ in Eq.~\eqref{eq:radial2s0}
\begin{equation}\label{FF_radial}
\dfrac{d^2R}{dy^2}+\dfrac{2}{y}\dfrac{dR}{dy}+\left[1+\dfrac{\epsilon}{y}-\dfrac{\gamma}{y^2}\right]R=0
\end{equation} 
with
\begin{eqnarray}
y&=&kr \nonumber\\
k^2&=&\omega^2-\mu^2\delta_{s,0} \nonumber\\ 
\epsilon&=&\dfrac{2is\omega-2\omega q Q+\left(2\omega^2-\mu^2\right)(1+a^2+Q^2)\delta_{n,0} }{k} \\
\gamma&=& \lambda^2-q^2Q^2-\omega(1+a^2+Q^2)\left[2qQ-\omega(1+a^2+Q^2)+2is\right]\delta_{n,0} \; .\nonumber
\end{eqnarray}
Note again that we are not studying the massive case for fermions. The $\delta_{s,0}$ factor in $k$ is emphasizing this -- it does not mean the $\mu^2$ term is absent for $s=1/2$. The general solution of Eq.~\eqref{FF_radial} is given in terms of Kummer functions
\begin{equation}\label{FF_Kummer}
R_{FF}=e^{-iy}y^{\sigma}\left[B_1M(u,v,2iy)+B_2U(u,v,2iy)\right] \; ,
\end{equation}
where
\begin{eqnarray}
\sigma&=&|s|-\frac{1}{2}+\sqrt{\left(|s|-\frac{1}{2}\right)^2+\gamma} \nonumber\\ 
 u&=& \sigma+1-|s|+i\frac{\epsilon}{2} \nonumber \\
v&=&2(\sigma+1-|s|) \; .
\end{eqnarray}
Eq.~\eqref{FF_Kummer} can be matched to the near horizon solution in the limit $y\ll 1$. This conditions implies $r\ll 1/k$,  so for consistency with the limit $r\gg 1$ we need $k$ small. The stretched form is
\begin{equation}\label{eq:stretchFF}
R_{FF}\rightarrow k^\sigma\left(B_1 r^\sigma+B_2\dfrac{\Gamma(v-1)}{\Gamma(u)}(2ik)^{1-v}r^{\sigma+1-v}\right) \; .
\end{equation}
It can be easily shown that within the same approximations as in Eq.~\eqref{eq:powers_match} the $r$-powers match with those in Eq.~\eqref{eq:stretchFF}. Then, up to an overall common constant 
\begin{eqnarray}
B_1&=&A_1 \nonumber \\
B_2&=&A_2\dfrac{\Gamma(u)(2ik)^{v-1}}{\Gamma(v-1)} \; .
\end{eqnarray}
Finally we expand in the far field limit $y\rightarrow +\infty$ to obtain (up to an overall common constant)
\begin{equation}\label{FF_asympt}
R_{FF}\rightarrow Y^{(in)}_s\dfrac{e^{-ikr}}{r^{1-|s|-s-i\varphi}}+Y^{(out)}_s\dfrac{e^{ikr}}{r^{1-|s|+s+i\varphi}} \; ,
\end{equation}
where
\begin{equation}
\varphi=\dfrac{\omega qQ}{k}-\dfrac{\left(\omega^2-\frac{\mu^2}{2}\right)(1+a^2+Q^2)}{k}\delta_{n,0} \; ,
\end{equation}
and
\begin{eqnarray}
Y^{(in)}_s&=&(2ik)^{-u}\left(\dfrac{B_1\Gamma(v)e^{i\pi u}}{\Gamma(v-u)}+B_2\right)\nonumber \\
Y^{(out)}_s&=&(2ik)^{u-v}\dfrac{B_1\Gamma(v)}{\Gamma(u)} \label{eq:LEapprox}\ .
\end{eqnarray}
Eq.~\eqref{FF_asympt} contains a combination of incoming and outgoing waves. However for the spin-half case the incoming/outgoing wave is dominant for $s=\pm 1/2$ respectively. Using the conserved number current it is possible to show~\cite{Creek:2007tw} that the transmission factor is 
\begin{equation}\label{eq:Transmission}
\mathbb{T}^{(4+n)}_{s,j,m}=1-\left|\dfrac{Y^{(out)}_{-|s|}}{Y^{(in)}_{|s|}}\right|^2 \; .
\end{equation}
For fermions, to find out the relative normalisation between $P_{1/2}$ and $P_{-1/2}$ we plug back the expansion~\eqref{FF_asympt} in the first order system~\eqref{eq:2s1radial}, equate order by order and obtain the relation
\begin{equation}\label{eq:Rel_norm}
Y^{(out)}_{-1/2}=\dfrac{2i\omega}{\lambda}Y^{(out)}_{1/2} \; .
\end{equation}
Since the relative normalisation between incoming and outgoing coefficients for the same $s$ is fixed, now we can insert Eq.~\eqref{eq:Rel_norm} in~\eqref{eq:Transmission} to obtain the transmission factor. For scalars, Eq.~\eqref{eq:Transmission} is also valid if we set $|s|=0$. 

\subsubsection{High energy approximation based on WKB arguments}
To complete the analytic picture, we present some arguments for a useful approximation in the high energy limit for scalars. This will give the leading asymptotic form for the transmission factors. 

The matching procedure in the previous section doesn't work in the high energy limit for two reasons. On one hand the powers in~\eqref{stretch_NH} and~\eqref{eq:stretchFF} no longer match at high energy, rotation, charges and masses. Secondly we are stretching the near horizon solution into $r\gg1$ and the far field solution into $r\ll 1/k$. If $k$ is large, then these conditions are incompatible and we are effectively stretching the far field solution too close to the horizon. 

To understand this problem we look into the WKB approximation for the scalar radial equation\footnote{Here we focus on the scalar case because the potential is real. A similar treatment can be applied to fermions using the method in Chandrasekhar's book~\cite{Chandrasekhar:1985kt} to reduce the complex potential to a real one.}. Some earlier works which have used the WKB approximation to compute transmission factors are~\cite{Iyer:1986np,Cho:2004wj,Cornell:2005ux,Grain:2006dg,Cho:2007de}. First note that the radial equation can be written in a Schr\"odinger-like form through a change of independent variable. Start by choosing
\begin{equation}\label{change_variable_x_y}
dy=\dfrac{dr}{\Delta} \; ,
\end{equation}
to obtain
\begin{equation}
\left(\dfrac{d^2}{dy^2}-V\right)R=0 \vspace{3mm} \; ,
\end{equation}
where $V:=\Delta U-K^2$ contains a leading term $-k^2r^4$ corresponding to the highest power of $r$ (all the other terms are suppressed). In the high energy limit, this term dominates the solution. In fact, we can formally write an infinite WKB series \cite{Adv_methods_WKB}
\begin{equation}
R\sim A_+\exp\left(k\sum_{n=0}^{\infty}\dfrac{S_n^+(y(r))}{k^n}\right)+A_-\exp\left(k\sum_{n=0}^{\infty}\dfrac{S_n^-(y(r))}{k^n}\right).
\end{equation}
It is easy to check~\cite{Adv_methods_WKB} that the leading correction reproduces the asymptotic form at infinity consistent with \eqref{FF_asympt}. A necessary condition for this approximation to be valid is
\begin{equation}\label{WKB_validity}
\left|\dfrac{dV}{dy}\right|\ll \left|V^{\frac{3}{2}}\right|\Leftrightarrow \left|\dfrac{dV}{dr}\Delta\right|\ll \left|V^{\frac{3}{2}}\right| \; ,
\end{equation}
which (to leading order in $r$) is just $r\gg 1/k$. This condition indicates that for large $k$ the field will start to take a WKB form not far from the horizon. Such result is not surprising if we note that these modes have very short wavelength, so the potential is almost constant along many wavelengths (except very close to the horizon). Furthermore, the WKB corrections obey
\begin{equation}\label{WKB_symm_coeffs}
S_n^{+}=\left\{ \begin{array}{ll}
-S_n^{-} &, n\ \mathrm{even} \vspace{2mm}\\
S_n^{-} &, n \ \mathrm{odd}
\end{array}\right.
\end{equation}
so the odd terms (which are purely real~\cite{Adv_methods_WKB}) only contribute with an overall common factor. As for the even terms, they are products of $\sqrt{V}$ times polynomial terms in $V$. In general there is an imaginary and a real part for each even order correction, but since our potential is real then it will either be real or imaginary. If $\sqrt{V}$ is real we get a relative change in amplitude between incoming and outgoing waves, whereas if it is imaginary the relative amplitude is fixed. But in the limit of $k$ large, the dominant term in the potential is $-k^2r^4$ which is negative so the square root is purely imaginary and the even order corrections only introduce a phase between incoming and outgoing waves. This means that in the region where the WKB solution is valid\footnote{This is the region connected to infinity such that $V<0$.}, the relative amplitude between incoming and outgoing modes stays fixed. The transmission coefficient can then be calculated at any point in such a region provided we have a suitable analytic expansion in terms of incoming and outgoing waves. Thus, in the high energy limit, the propagation of the field along a thin region outside the horizon, determines the behaviour of the greybody factors.

This behaviour can be seen explicitly in \eqref{stretch_NH}. There the scalar $r$-powers have a common factor $r^{-(n+1)(1-B_*/2)}$ multiplied by 
\begin{equation}\label{NH_WKB_powers}
r^{\pm\sqrt{(1-B_*)^2/4-p^2+D_*}} \; . 
\end{equation}
In the high energy limit the argument of the square root in Eq.~\eqref{NH_WKB_powers} becomes negative and we obtain a relative phase between the two modes which are respectively outgoing and incoming. The transmission coefficient follows under the single approximation $k\gg 1$
\begin{equation}\label{transmission_largek}
\mathbb{T}^{4+n}_{0,j,m}=1-\left|\dfrac{A_+}{A_-}\right|^2=1-\left|\dfrac{A_1}{A_2}\right|^2 \; .
\end{equation}

\section{Results}\label{sec:Results}
In this section we plot various quantities, using the approximations developed in Sect.~\ref{sec:Evaporation}. The physically most relevant are those in \eqref{eq-flux-spectrum}, \eqref{eq-power-spectrum}, \eqref{eq-angular-spectrum} and \eqref{eq-charge-spectrum}. When integrated over $\omega$ and summed over particle type they give the rates of emission of particle number, energy, angular momentum and charge. Nevertheless we still plot the transmission factors to keep track of where the new effects enter. We discuss scalars and fermions in parallel whenever possible and present the effects of particle mass and charge separately. We do not present plots with rotation to avoid repetition of results which have been studied (without mass and charge) in previous publications~\cite{Ida:2002ez,Ida:2005ax,Duffy:2005ns,Casals:2005sa,Ida:2006tf,Casals:2006xp}. However we have checked that within our approximations, our general result agrees with those special cases. 

\subsection{The effect of particle mass}
\begin{figure}[t]
\begin{center}
\includegraphics[scale=0.65,clip=true,trim=2cm 0cm 0cm 0cm]{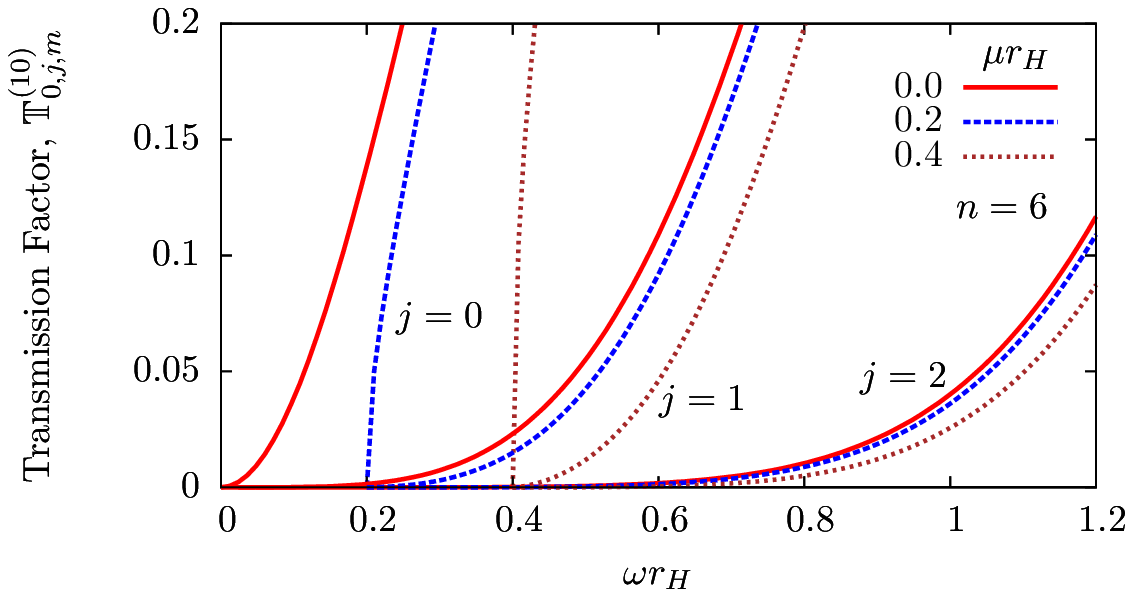}
\includegraphics[scale=0.65,clip=true,trim=2cm 0cm 0cm 0cm]{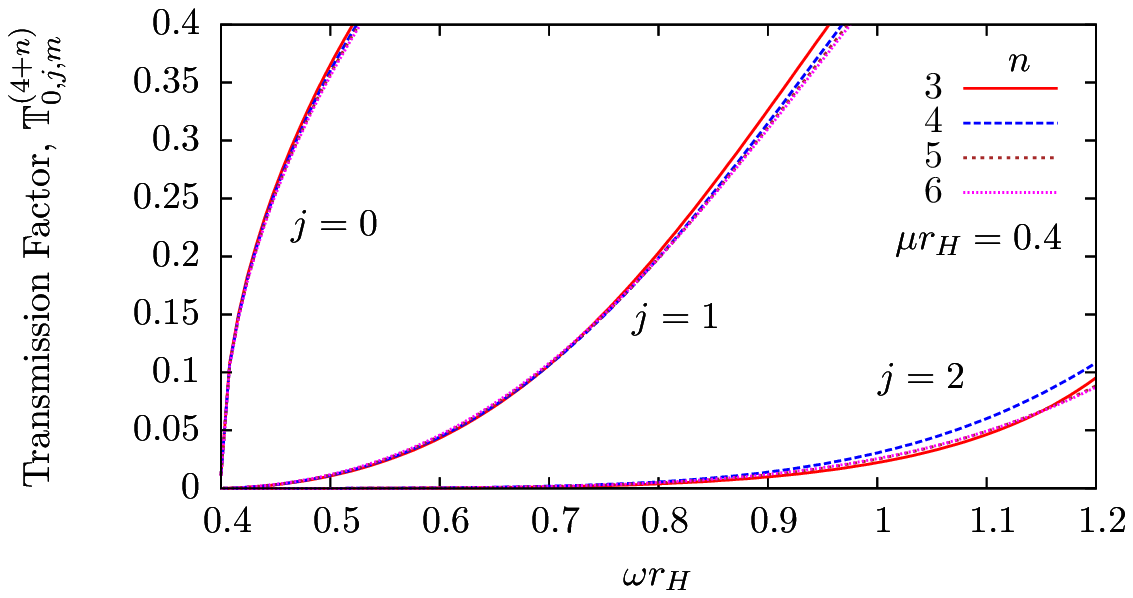}
\includegraphics[scale=0.65,clip=true,trim=0.6cm 0cm 0cm 0cm]{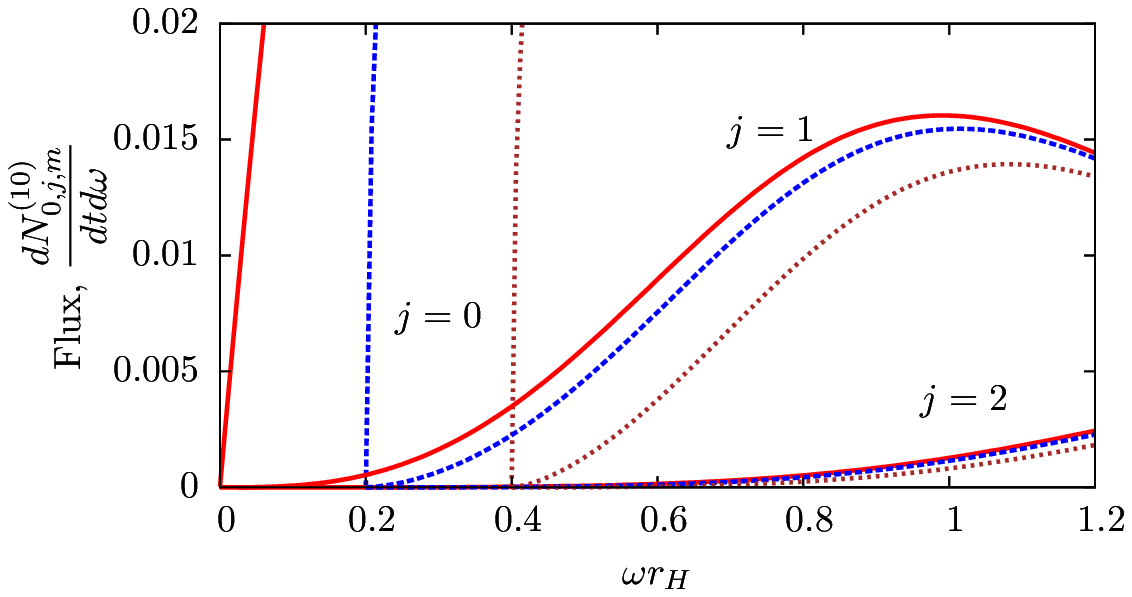} 
\includegraphics[scale=0.65,clip=true,trim=0.6cm 0cm 0cm 0cm]{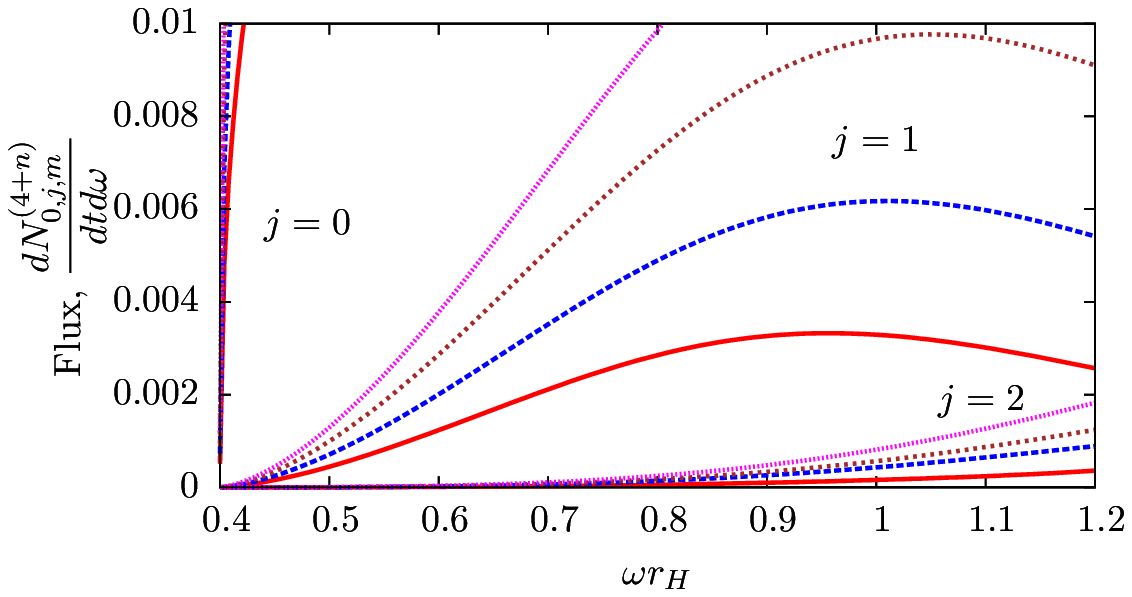}
\end{center}
\caption{\emph{Scalar transmission factors and fluxes for $n = 6$ (left) and variable $n$ (right).} The left plots show variation with particle mass $\mu$ in natural units $r_H^{-1}$ for $n=6$ and the right plots show variation with $n$ for $\mu$ fixed. The top plots show the transmission factors $\mathbb{T}^{(4+n)}_k$ and the bottom plots show number fluxes, for a range of $j$ modes. The curves are naturally grouped by $j$, rotation is off and the line colour/type is the same for top and bottom plots.}
\label{fig-2s0mass-variable}
\end{figure}
Current modelling of black hole evaporation in ($4+n$)-dimensional TeV gravity scenarios does not take into account the dynamical consequences of non-zero mass for the emitted particles. This effect is important if the energy of the particle emitted during the evaporation is close to its mass. For Standard Model heavy particles, the top quark ($m_t\sim 170 \ \mathrm{GeV}$) the Z ($m_Z\sim 91 \ \mathrm{GeV}$), the W ($m_W\sim 80 \ \mathrm{GeV}$) and the Higgs boson, the effect will not be negligible.  

In Fig.~\ref{fig-2s0mass-variable} we present some representative curves for the transmission factor and the number flux. As mentioned before the approximation becomes better with larger $n$ so most of our plots will be for $n=6$, except for when we focus on the $n$ dependence where we use $n\geq 3$. 

The most prominent property of the left-hand-side plots is a smooth drop close to the mass $\mu$. The higher the partial wave the less steep this is but however there is always a horizontal shift (see for example the $j=1$ mode). This effect is quite important close to the mass threshold where the probability of emission is suppressed. This is in contrast with the simplified approach in current BH event generators where the spectrum is cut off sharply at $\omega=\mu$. 

Furthermore, for example the number flux for the $j=1$ mode shows how increasing the mass of the particle not only suppresses the flux around $\omega\sim \mu$ but also the total area under the curve. Massive particles are therefore less likely to be produced. This effect was previously studied in four dimensions, for example numerically, in Page's paper for leptons~\cite{Page:1977um}.

The right hand side plots show how the transmission factor is very mildly dependent on $n$ (at least in the limit of small $\mu$). However the flux plot displays a strong variation with $n$ which is due to the strong dependence of the Hawking temperature appearing in the thermal factor. 

\subsection{The effect of BH charge on neutral particles}
\begin{figure}[t]
\begin{center}
\includegraphics[scale=0.65,clip=true,trim=2cm 0cm 0cm 0cm]{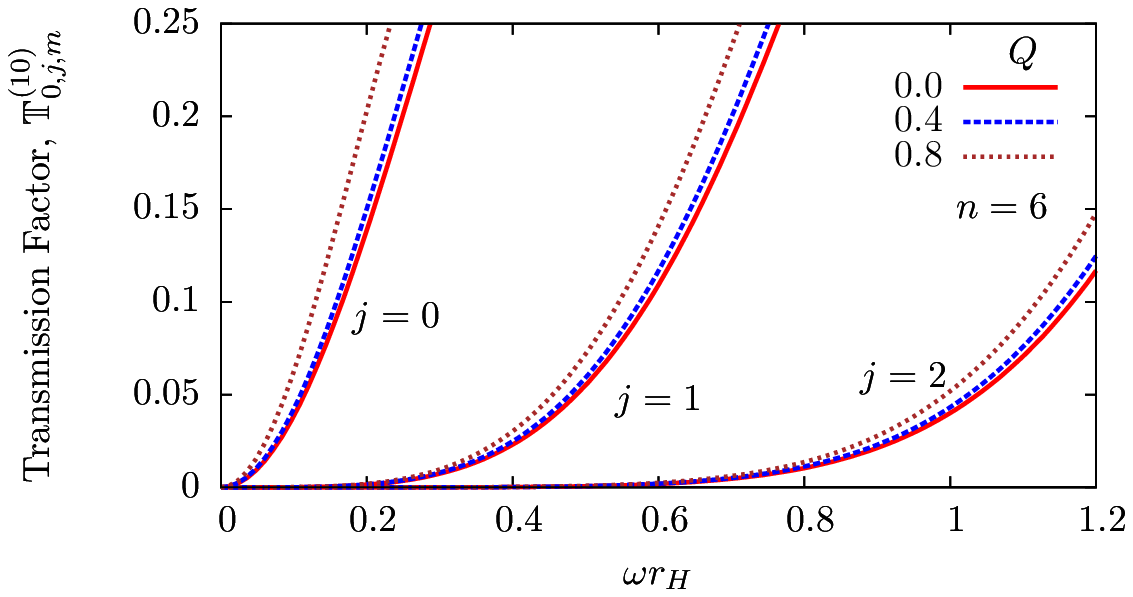}
\includegraphics[scale=0.65,clip=true,trim=2cm 0cm 0cm 0cm]{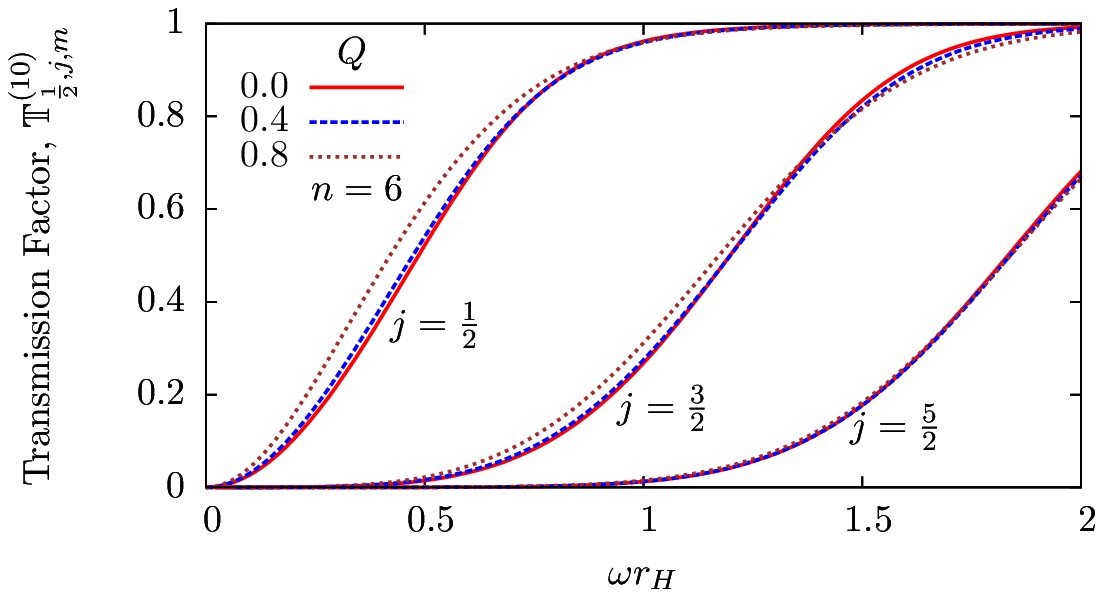}
\includegraphics[scale=0.65,clip=true,trim=0.6cm 0cm 0cm 0cm]{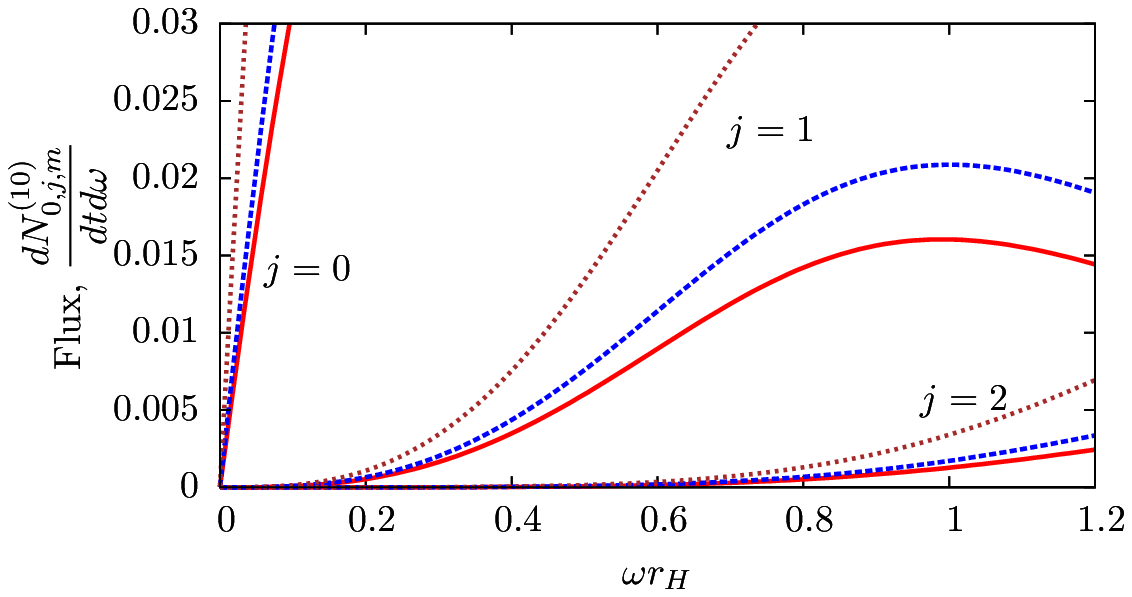} 
\includegraphics[scale=0.65,clip=true,trim=0.6cm 0cm 0cm 0cm]{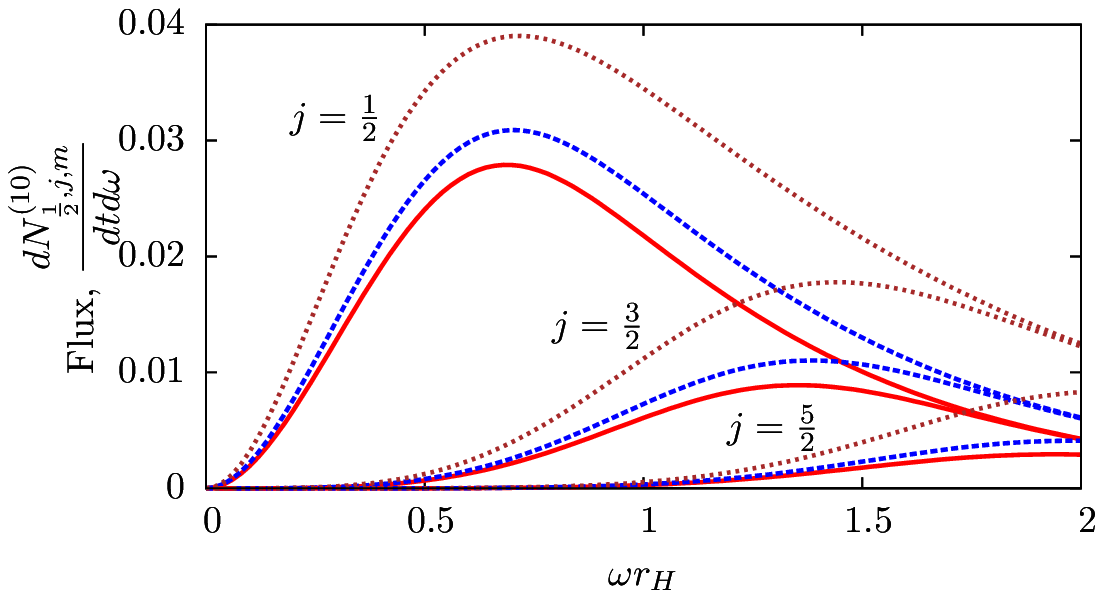}
\end{center}
\caption{\emph{Transmission factors and fluxes for neutral scalars and fermions.} The left plots show spin $0$ and the right plots show spin $1/2$. The top plots show transmission factors $\mathbb{T}_k$ and the bottom plots show number fluxes, for a range of $j$ modes and different black hole charges $Q$. Rotation is off, $\mu=0$ and the line colour/type is the same for top and bottom plots.}
\label{fig-trans-fluxes-neutral}
\end{figure} 
The next effect we consider is black hole charge. Neutral particles simply feel a different gravitational field around the black hole. So by studying neutral particles we disentangle the gravitational effect from the electromagnetic effect (since $q=0$). 

 In Fig.~\ref{fig-trans-fluxes-neutral} we present plots for transmission factors and fluxes. Here we focus on $n=6$ for scalars and fermions. We should note that some of the plots for fermions will display extrapolated results beyond the small energy limit. This turns out to be quite well behaved, which is due to the better matching of $r$-powers as pointed out in Eq.~\eqref{eq:condition_ell}. 

From the gravitational point of view, the main effect of $Q$ is to decrease the horizon radius and consequently increase the Hawking temperature. This is clearly seen in the transmission factors of Fig.~\ref{fig-trans-fluxes-neutral}, where all the curves are pushed up with increasing $Q$. The same happens with the fluxes where the effect is even larger, due to the strong dependence of the thermal factor on $Q$ through the Hawking temperature -- see Eq.~\eqref{eq-flux-spectrum}.

\subsection{The effect of particle charge} 
For particles with non-zero charge, in addition, we have a Coulomb repulsion/attraction  according to whether the particle has same/opposite sign charge compared to the black hole. For definiteness we take the black hole charge to be positive.
\begin{figure}[t]
\begin{center}
\includegraphics[scale=0.65,clip=true,trim=2cm 0mm 0mm 0cm]{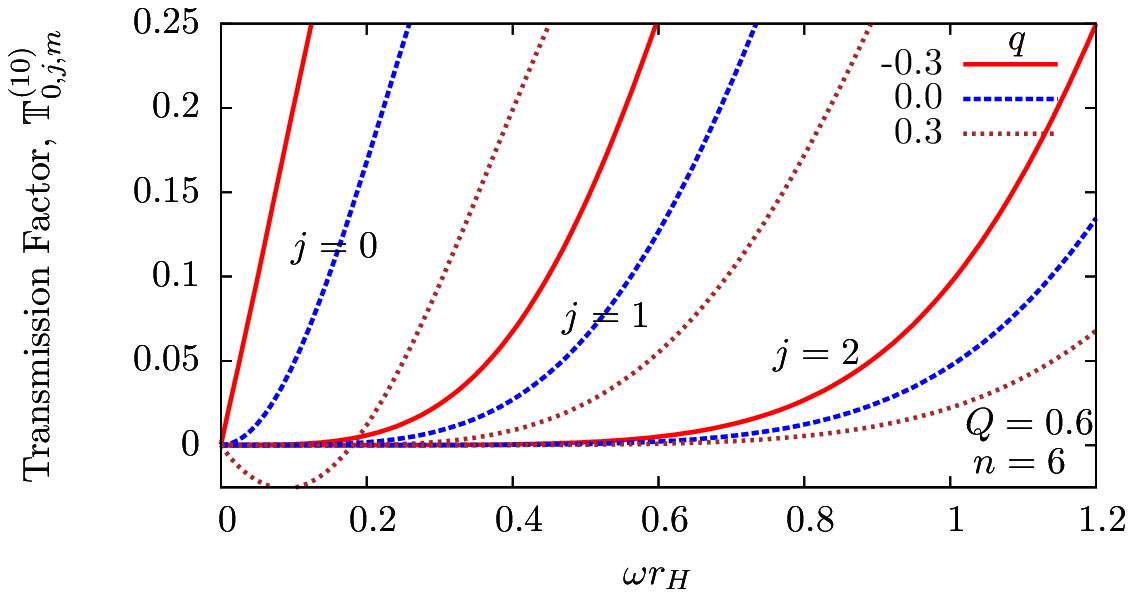}
\includegraphics[scale=0.65,clip=true,trim=2cm 0cm 0cm 0cm]{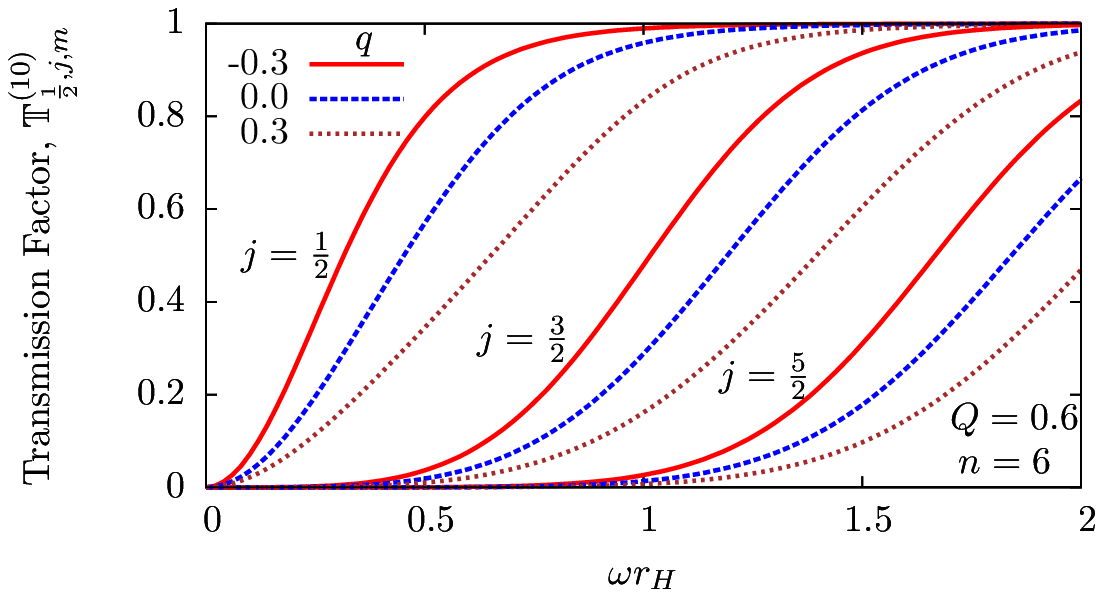}
\includegraphics[scale=0.65,clip=true,trim=0.6cm 0cm 0cm 0cm]{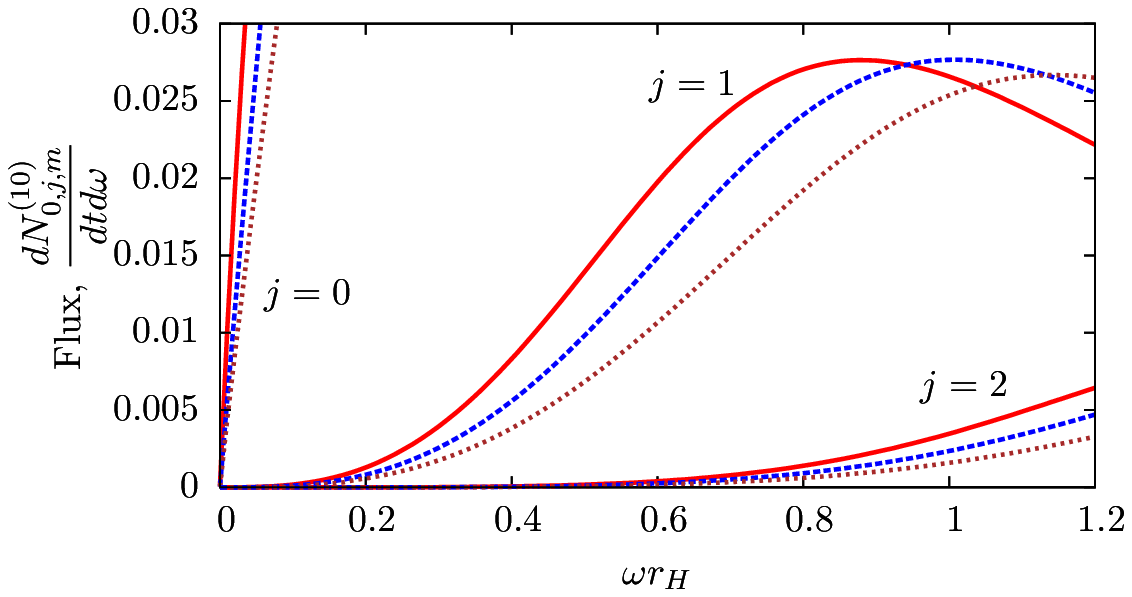} 
\includegraphics[scale=0.65,clip=true,trim=0.6cm 0cm 0cm 0cm]{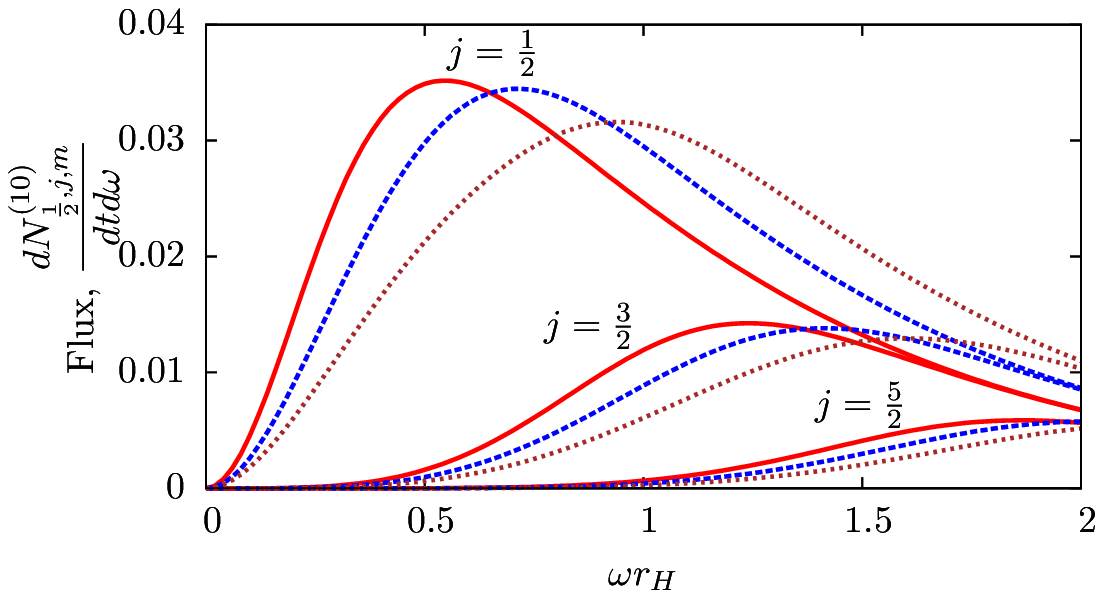}
\end{center}
\caption{\emph{Transmission factors and fluxes for charged scalars and fermions.} The left plots show spin $0$ and the right plots show spin $1/2$. The top plots show transmission factors $\mathbb{T}_k$ and the bottom plots show number fluxes, for a range of $j$ modes and different particle charge $q$ with $Q=0.6$. Rotation is off, $\mu=0$ and the line colour/type is the same for top and bottom plots.}
\label{fig-trans-fluxes-charged}
\end{figure} 

\begin{figure}[t]
\begin{center}
\includegraphics[scale=0.65,clip=true,trim=2cm 0cm 0cm 0cm]{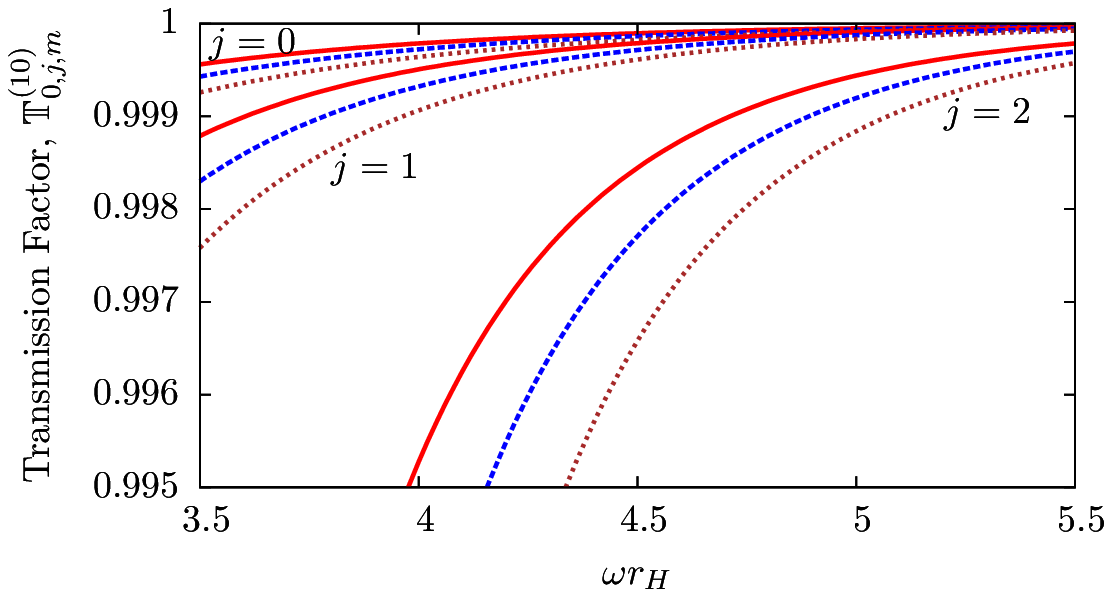}
\includegraphics[scale=0.65,clip=true,trim=0.6cm 0cm 0cm 0cm]{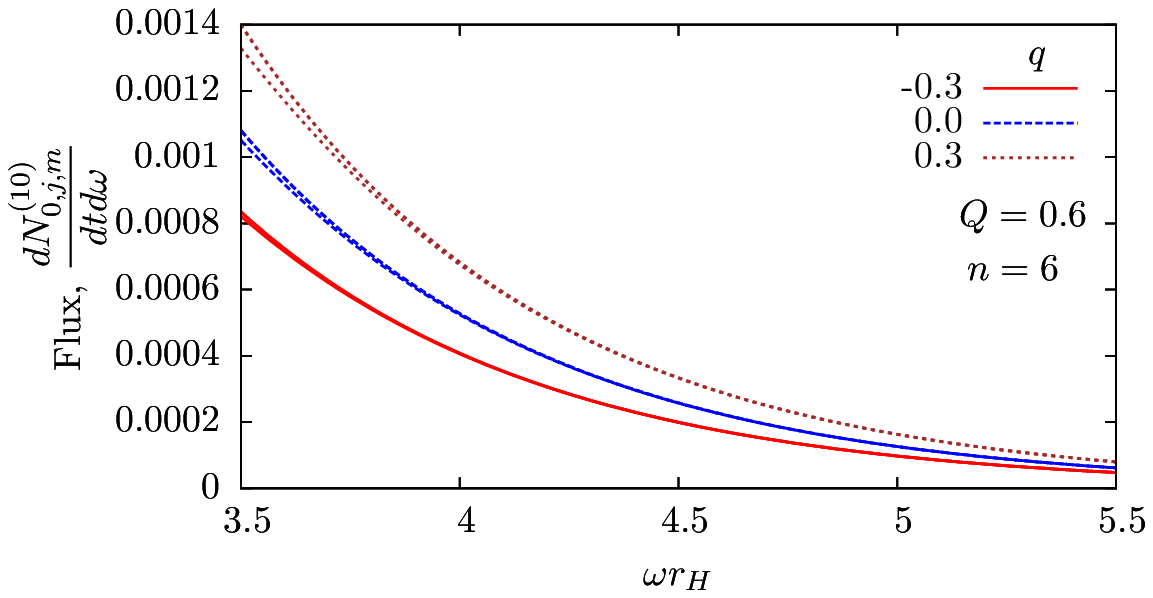} 
\end{center}
\caption{\emph{Asymptotic high energy transmission factors and fluxes for charged scalars.} The left plot shows transmission factors $\mathbb{T}^{(4+n)}_k$ and the right plot shows fluxes for a range of $j$ modes. Rotation is off, $\mu=0$ and the line colour/type is the same for both plots.}
\label{fig-trans-fluxes-chargedHE}
\end{figure}
In Fig.~\ref{fig-trans-fluxes-charged} we plot transmission factors and fluxes for scalars and fermions, for various charges, $n=6$ and $Q=0.6$. 
It is important to note here that the Coulomb type coupling appearing in the radial equation is \[qQ=(\sqrt{\alpha}z)(\sqrt{\alpha}Z)\simeq (0.1z)(0.1Z) \; .\] For an LHC black hole, at production, $|Z|\leq 4/3$ and $|z|\leq 1$. So the figures we have chosen are above their typical values. However it is easier to see the differences in the curves. Furthermore there may be stages during the evaporation where the black hole charges up so this region of parameters is not completely unphysical. 

The main features of Fig.~\ref{fig-trans-fluxes-charged} are as follows. For scalars we can see clearly the phenomenon of superradiance in the top plot for particles with the same charge as the black hole, where $\mathbb{T}^{(10)}_k<0$. However, this does not favour the emission of positively charged particles because the negative charge transmission factors are greatly enhanced. This is clear in the flux plot where all the curves at low energies are higher for negative charge. This can be understood physically by recalling that the transmission factor describes the probability of a wave incident from infinity to be transmitted down the black hole. Since negatively charged particles are attracted by the Coulomb potential and positively charged particles are repelled, we would expect negative charges to have higher transmission factors. This is confirmed for fermions in a wider range of energies. The other main feature is that at higher energies, the thermal factor (which favours discharge) dominates and the tendency is inverted, i.e. positively charged particles are favoured.    
This is confirmed for the scalar case in the high energy limit in Fig.~\ref{fig-trans-fluxes-chargedHE}, where the transmissions factors are still larger for negatively charged particles, but however, since they are close to their asymptotic value $\mathbb{T}^{(10)}_k=1$ the thermal factor dominates.    

\begin{figure}[t]
\begin{center}
\includegraphics[scale=0.65,clip=true,trim=1.8cm 0cm 0mm 0cm]{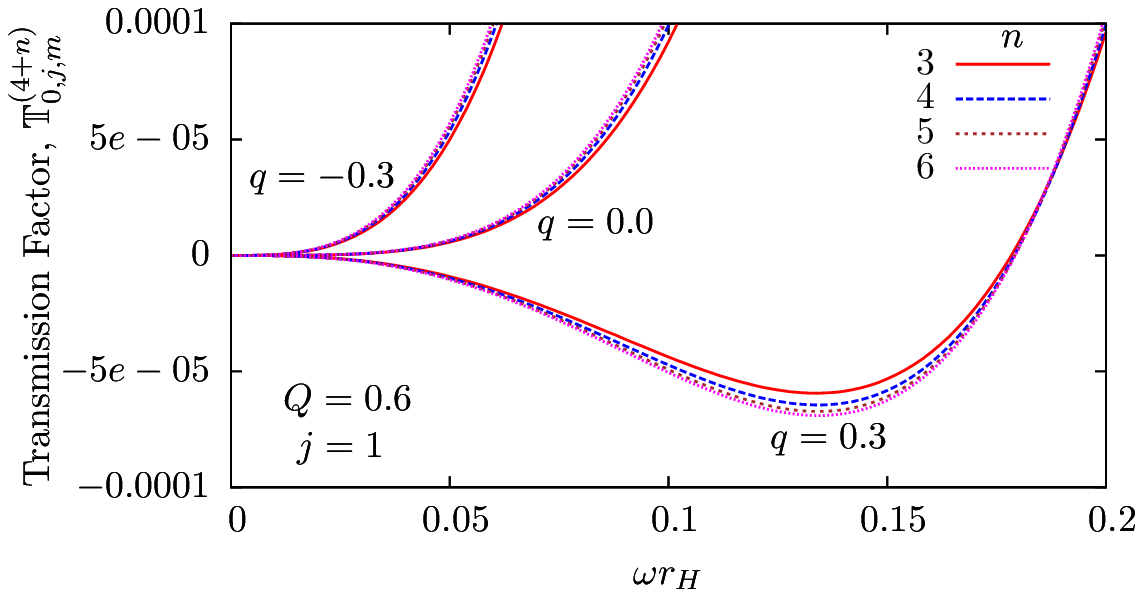}
\includegraphics[scale=0.65,clip=true,trim=1.7cm 0cm 0cm 0cm]{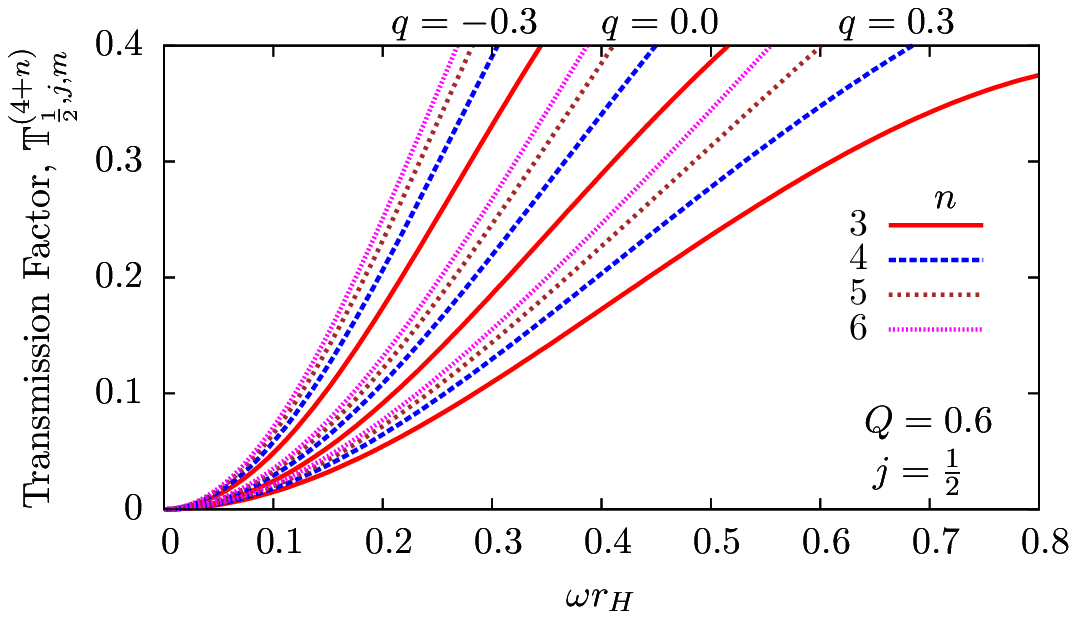}
\includegraphics[scale=0.65,clip=true,trim=0.3cm 0cm 0mm 0cm]{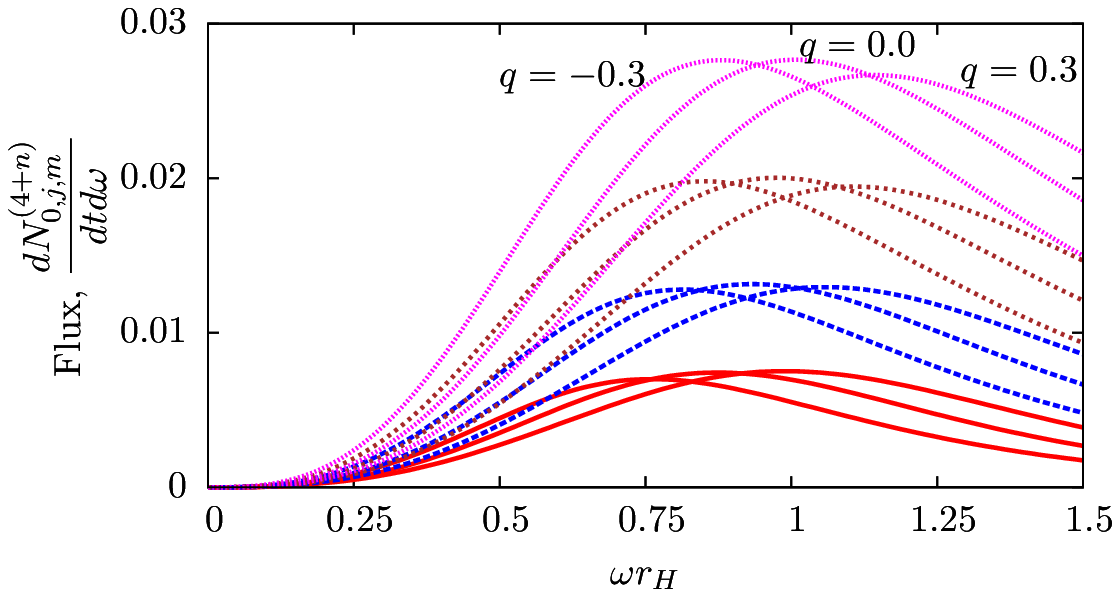} 
\includegraphics[scale=0.65,clip=true,trim=0.7cm 0cm 0mm 0cm]{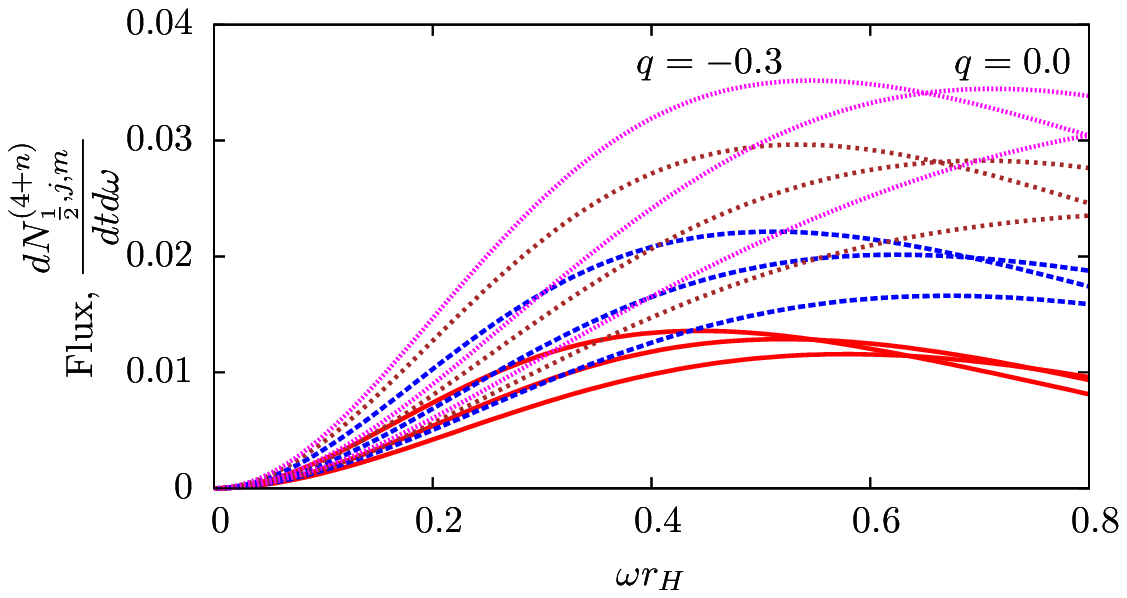}
\end{center}
\caption{\emph{Transmission factors and fluxes for charged scalars and fermions.} The left plots show spin $0$ and the right plots show spin $1/2$. The top plots show transmission factors $\mathbb{T}_k$ and the bottom plots show number fluxes, for a range of charges $q$ and various numbers of extra dimensions $n$ with $Q=0.6$ and $j=1$ for scalar and $j=1/2$ for fermions. Rotation is off, $\mu=0$ and the line colour/type is the same for top and bottom plots.}
\label{fig-trans-fluxes-nVar}
\end{figure}
Finally, Fig.~\ref{fig-trans-fluxes-nVar} shows the variation with $n$. Here the transmission factors for scalars are weakly dependent on $n$ whereas for fermions we have a stronger effect. This is due to extra $n$ dependent factors in the wave equation as for example the term $2|s|/A\sim 1/(n+1)$ in $B$ -- Eq.~\eqref{NH_approximations}. For the fluxes the separation is larger due to their stronger $n$ dependence through the Hawking temperature. In general, similarly to neutral black holes, the effect of $n$ is to increase the total fluxes.

\section{Conclusions}\label{sec:Conclusions}

We have presented a calculation of transmission factors for brane-charged massive scalars and charged massless fermions on a $(4+n)$-dimensional brane-charged rotating black hole in the low and high energy limits. Our main theoretical results are:
\begin{enumerate}
\item
  The construction of an approximate background with a stationary axisymmetric gravitational field and brane-electromagnetic field -- Eq.~\eqref{KN_ansatz} and~\eqref{ansatz_metric}.
\item
  The derivation of separated scalar and fermion wave equations for particles with non-zero mass $\mu$ and charge $q$ -- Eqs.~\eqref{eq:radial2s0},\eqref{eq:angular2s0},~\eqref{eq:2s1radial} and~\eqref{eq:2s1angular}. The radial and angular equations we obtained can be integrated for arbitrary parameters given suitable boundary conditions.
\item
  The application of a well known procedure in the low energy limit to obtain transmission factors (and consequently Hawking fluxes), for the new cases of: brane charged massive scalar field; and massless charged fermion field -- Eq.~\eqref{eq:Transmission} and \eqref{eq:LEapprox}. Furthermore we used the WKB approximation to find the asymptotic behaviour for the scalar case -- Eq.~\eqref{transmission_largek}.
\end{enumerate}

Finally the numerical evaluation of the approximate transmission factors showed important new features which are relevant for LHC phenomenology and the development of BH event generators. The two central results are:
\begin{itemize}
\item 
  For massive particles, our scalar analysis shows a damping of the spectrum close to the threshold $\omega \simeq \mu$ as well as an overall reduction of the area under the flux curves -- Fig.~\ref{fig-2s0mass-variable}. The main consequence for LHC phenomenology is that production of massive particles such as the top, W, Z and Higgs boson (which have masses of the same order of magnitude as the typical $1/r_H\sim 100 \ \mathrm{GeV}^{-1}$) is highly suppressed at low energies.  
\item
  Black hole discharge is subdominant -- Eq.~\eqref{schwinger_estimate}, Figs.~\ref{fig-trans-fluxes-charged} and~\ref{fig-trans-fluxes-chargedHE}. This is another important point for LHC phenomenology and the development of event generators which tend to enforce quick discharge. Nevertheless, black hole events at the LHC will have non-zero charge, so statistically we would expect a fraction of them to charge up. For charged black holes our plots show that the flux spectra for positive and negative charges are split. Thus negatively charged particles are biased towards low energies whereas positively charged particles are biased towards higher energies. So the dynamical model of discharge should still be incorporated since it will produce an asymmetry in the energy spectrum of positive/negative charged particles.    
\end{itemize}

To summarize, the effects of mass and charge are important for improving the modelling of black hole events from high energy collisions in large extra dimensions scenarios, and may provide further signatures of black hole events such as charge asymmetries.  Two points we haven't discussed which deserve further attention are those of QCD charges and the possible restoration of electroweak symmetry close to the black hole. Both can be treated using an improved model based on the ideas we have discussed.  


\section*{Acknowledgements}

MS thanks colleagues in the Cambridge SUSY Working Group for helpful
discussions. This work was supported by the FCT grant SFRH/BD/23052/2005.

\bibliography{DraftCharge}
\bibliographystyle{JHEP}

\end{document}